\documentclass[iicol,pdflatex,sn-mathphys-ay]{sn-jnl}

\usepackage{graphicx}%
\usepackage{multirow}%
\usepackage{amsmath,amssymb,amsfonts}%
\usepackage{amsthm}%
\usepackage{mathrsfs}%
\usepackage[title]{appendix}%
\usepackage{xcolor}%
\usepackage{textcomp}%
\usepackage{manyfoot}%
\usepackage{booktabs}%
\usepackage{algorithm}%
\usepackage{algorithmicx}%
\usepackage{algpseudocode}%
\usepackage{listings}%
\usepackage{comment}
\usepackage[normalem]{ulem}
\usepackage{xspace} 
\usepackage{subfigure}
\usepackage{bm}
\let\cite\citep
\usepackage[table]{xcolor}

\newcommand{\singlefocal}{fixed-focal\xspace}

\newif\ifchange

\changefalse

\ifchange
\newcommand{\additional}[1]{\textcolor{teal}{#1}}
\newcommand{\deleted}[1]{\textcolor{orange}{\sout{#1}}}

\else
\newcommand{\additional}[1]{\textcolor{black}{#1}}

\newcommand{\deleted}[1]{}
\fi

\theoremstyle{thmstyleone}%
\theoremstyle{thmstyletwo}%

\theoremstyle{thmstylethree}%

\begin{document}

\title[Article Title]{Varifocal Displays Reduce the Impact of the Vergence-Accommodation Conflict on 3D Pointing Performance in Augmented Reality Systems}


\author*[1]{\fnm{Xiaodan} \sur{Hu}}\email{x-hu@shibaura-it.ac.jp}

\author[2]{\fnm{Monica} \sur{Perusquía-Hernández}}\email{perusquia@ieee.org}

\author[3]{\fnm{Mayra} \spfx{Donaji} \sur{Barrera Machuca}}\email{mbarrera@ucalgary.ca}

\author[4]{\fnm{Anil} \spfx{Ufuk} \sur{Batmaz}}\email{ufuk.batmaz@concordia.ca}

\author[5]{\fnm{Yan} \sur{Zhang}}\email{yan-zh@sjtu.edu.cn}

\author[6]{\fnm{Wolfgang} \sur{Stuerzlinger}}\email{w.s@sfu.ca}

\author*[2]{\fnm{Kiyoshi} \sur{Kiyokawa}}\email{kiyo@is.naist.jp}

\affil*[1]{
\orgname{Shibaura Institute of Technology}, \country{Japan}}
\affil[2]{
\orgname{Nara Institute of Science and Technology}, \country{Japan}}
\affil[3]{
\orgname{University of Calgary}, \country{Canada}}
\affil[4]{
\orgname{Concordia University}, \country{Canada}}
\affil[5]{
\orgname{Shanghai Jiao Tong University}, \country{China}}
\affil[6]{
\orgname{Simon Fraser University}, \country{Canada}}


\abstract{
This paper investigates whether a custom varifocal display can improve 3D pointing performance in augmented reality (AR), where the vergence-accommodation conflict (VAC) is known to impair interaction. Varifocal displays have been hypothesized to alleviate the VAC by dynamically matching the focal distance to the user’s gaze-defined target depth. Following prior work, we conducted a within-subject study with 24 participants performing an ISO 9241-411 pointing task under varifocal and fixed-focal viewing. Overall, varifocal viewing yielded significantly higher performance than the fixed-focal baseline across key interaction metrics, although the magnitude and even the direction of the benefit varied across individuals. In particular, participants' responses exhibited a baseline-dependent pattern, with smaller improvements (or occasional degradation) observed for those with better baseline performance. Our findings suggest that varifocal technology can improve AR pointing performance relative to fixed-focal viewing, while highlighting substantial individual differences that should be considered in design and evaluation.}

\keywords{3D pointing, Fitts' Law, AR, Vergence-Accommodation Conflict, Varifocal Display.}



\maketitle

\section{Introduction}

\begin{figure*}
    \centering
    \includegraphics[width=\linewidth]{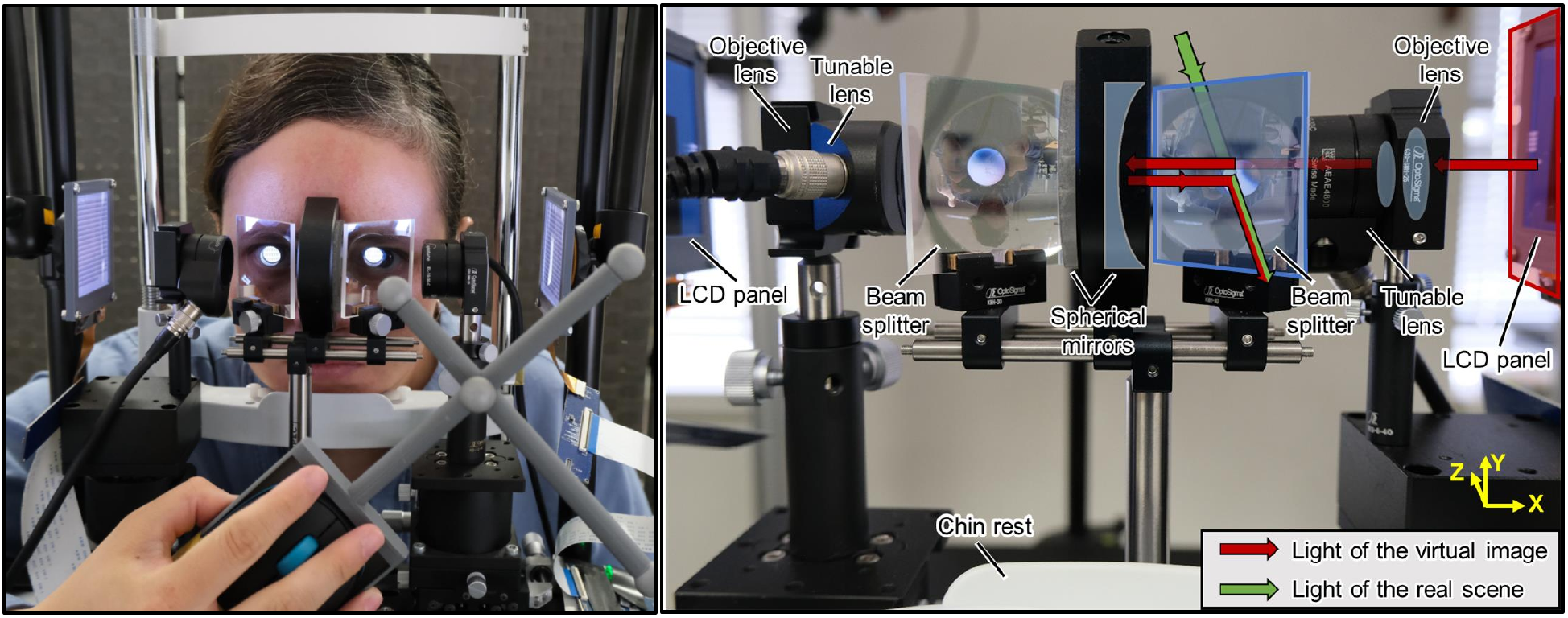}
    \caption{The custom varifocal display. (Left) A user pointing at a virtual target with the wand input device, as viewed through the varifocal display. (Right) The components of the varifocal display. Green arrows represent light from the real scene, while red arrows represent light from the virtual image, reflected by the spherical mirror and beamsplitter. Both converge at the beamsplitter and enter the human eye together.}
    \label{fig:teaser}
\end{figure*}

Thanks to recent technological advances, modern augmented reality (AR) head-mounted displays (HMDs) feature low latency in rendering and tracking, as well as a wide field-of-view (FOV) that enables seamless viewing~\cite{Koulieris19Near}. Modern AR HMDs also adjust the distance between the lenses to match the user's inter-pupillary distance (IPD), i.e., the distance between the centers of the eyes' pupils, supporting accurate stereopsis. Some modern HMDs integrate eye-tracking technology that detects where the user is looking.  
Despite these advancements, previous work identified that modern AR HMDs still do not display all depth cues correctly~\cite{Armbruster2008a, Lin2014a, Renner2013c, Singh2010, 9756824, 8985328}.

Depth perception issues are influenced by both human visual system limitations, such as age-related near-field problems, stereo deficiencies, or diplopia~\cite{Bruder2013a, Hess2015a, Zaroff2003a}, and technical constraints of display systems~\cite{Singh2018, 10.1145/3611659.3615686}, such as screen distance~\cite{8985328} and image elements~\cite{9756824}. Another issue that affects depth perception is the \emph{vergence-accommodation conflict} (VAC)~\cite{10.1145/3611659.3615686}, which arises from the way HMDs present stereo content. AR HMDs project two different images to the user's eyes, with the two viewpoints corresponding to the eyes' positions. Each image is displayed or projected onto a fixed plane by the HMD, typically via a beam splitter and AR lenses. Thus, when displaying 3D content at a depth different from the fixed focal plane, the user's eyes experience a mismatch between focusing on the display plane (accommodation) and rotating the eyes to perceive the object at its correct visual depth (vergence). The VAC impairs pointing performance in stereo systems~\cite{batmaz2019:DoStereoDeficinciesAffesct, 10.1145/3611659.3615686, Wang24} and large stereo displays~\cite{barreramachucaEffectStereoDisplay2019a}.

Researchers have hypothesized that one way to address the VAC is to correct the focal plane in real-time, for example, by actively shifting the location of a single (or multiple) display planes based on the output of a gaze tracker, e.g., in a varifocal stereo display~\cite{7226865, liuNovelPrototypeOptical2010}. Varifocal stereo displays use lenses with variable focal lengths that dynamically adjust focus based on the user's gaze position~\cite{changMultifocalDisplaysDense2018}. Previous work has presented varifocal stereo displays that reduce the accommodative mismatch associated with the VAC and provide correct visual depth cues~\cite{Suyama_2000, padmanaban2017optimizing}. However, these studies only evaluated their effect on the visual \emph{perception} of 3D targets and not on the user \emph{interaction} with such targets.

In this paper, we aim to investigate whether varifocal stereo displays can eliminate the interaction performance disadvantage introduced by the VAC inherent in most current stereo display technologies. While prior research has primarily focused on the perceptual effects of the VAC, the impact on 3D interaction remains underexplored, with most work evaluating only single-focal displays~\cite{batmaz2019:DoStereoDeficinciesAffesct, 10.1145/3611659.3615686, barreramachucaEffectStereoDisplay2019a, Batmaz:2022:VRST}. 
Understanding the effect of the VAC on interaction in varifocal stereo displays is important, as this technology represents a potential direction for future AR HMDs. 
At the same time, human visual perception and visuomotor processing are known to vary across individuals, raising the question of whether the interaction benefits of varifocal support are uniformly expressed across users~\cite{hu2024perception, drey2023investigating}.

We employed a custom-built AR varifocal display, whose design had been previously validated through perceptual studies~\cite{liuNovelPrototypeOptical2010}. We assessed interaction performance using a Fitts’ law task~\cite{fitts1954information} conforming to the ISO 9241-411 standard~\cite{ISO2015} to isolate the effects of depth on interaction involving target acquisition in two movement directions—one with significant depth variation and one without. By leveraging a validated optical system design and an established performance evaluation protocol, we quantify the influence of the VAC by comparing pointing accuracy and speed across varying depth planes using a custom-made stereo display that offers both varifocal and fixed-focal display modes in AR (\autoref{fig:teaser}). Moreover, as the physical setup can affect overall task performance (e.g., posture and reach)~\cite{brogmus1991effects, rohr2006upper}, which can potentially mask display-related effects, we repeated the same protocol with a new participant group using a prototype with revised ergonomics, enabling us to verify that the observed trends are not an artifact of an overly constrained setup.

Our work is the first experiment on virtual hand interaction using an AR varifocal display. It still builds on previous research, which identified that stereo display deficiencies affect interaction~\cite{batmaz2019:DoStereoDeficinciesAffesct, 10.1145/3611659.3615686, barreramachucaEffectStereoDisplay2019a, batmaz2023virtualhandvac, Batmaz:2022:VRST}. It also extends research on the effect of the VAC on interaction~\cite{Batmaz:2022:VAC}. Our contributions are:

\begin{itemize}
\item \textbf{A varifocal AR display tailored for evaluating mid-air interaction:}  
Building on insights from prior VAC studies~\cite{Batmaz:2022:VAC}, we constructed a stereo AR varifocal display based on a validated design~\cite{liuNovelPrototypeOptical2010} and optimized it for virtual hand pointing.

\item \textbf{A study of interaction with an AR varifocal display:}
We compare \singlefocal and varifocal modes and report performance outcomes across two study iterations, including a revised ergonomic setup to reduce potential performance constraints of the prototype. Results reveal interaction performance patterns consistent with VAC mitigation under varifocal viewing.

\item \textbf{Baseline-dependent effects of varifocal viewing:}
We observe that participants with better baseline performance under the fixed-focal condition tend to show smaller improvements under varifocal viewing, and in some cases even degraded performance. Other participants exhibited larger improvements in the varifocal condition.

\end{itemize}

\section{Literature Review}
We first review prior research on the VAC and its effect on interaction performance. Then, we cover stereo display technologies designed to address the VAC, followed by studies on depth perception in VE using stereo displays. Lastly, we examine the literature on virtual hand-based 3D object selection.

\subsection{VAC and Interaction Performance}
Past research has extensively documented the negative effects of \additional{a change in depth} on user interaction performance. For instance, Barrera Machuca and Stuerzlinger~\cite{barreramachucaEffectStereoDisplay2019a} observed that movements along the line of sight are 25\% slower than lateral movements in stereo displays due to a secondary sub-movement, which is not observed with physical targets. They identified this secondary sub-movement as a depth-estimation error that extends the correction phase of the hand movement\additional{, and speculated that display-related depth cue limitations may contribute to this effect}. Yet, they used a singlefocal stereo display rather than a varifocal one.

\additional{Follow-up work by }Batmaz et al.~\cite{Batmaz:2022:VAC} showed that the VAC significantly reduces 3D selection performance and adversely affects virtual hand interaction if targets are at varying depths. 
For the user study, they used a custom-made, benchtop-sized multifocal stereo display with only three depths, which limited the applicability of their results for potential commercial realizations.
Although prior work speculated that varifocal stereo displays could reduce the effects of VAC on interaction, no studies have quantitatively evaluated 3D pointing performance using such displays. Thus, the advantages of varifocal displays over other stereo displays for interaction remain unclear.

\subsection{Stereo Display Systems That Address the VAC}

The human visual system relies on depth cues such as accommodation, which enables the eye to focus on objects at varying distances. Most AR HMDs fail to simulate these cues accurately, resulting in the VAC, which leads to discomfort and reduces immersion~\cite{kramidaResolvingVergenceAccommodationConflict2016a}. Varifocal and multifocal displays have emerged as promising solutions, with recent developments focusing on more compact and efficient designs.

Most commercial HMDs, such as the Oculus Quest 3\footnote{https://www.meta.com/quest/quest-3} 
and the HTC VIVE Pro 2\footnote{https://www.vive.com/us/product/vive-pro2}, utilize fixed focal planes set at specific distances, typically beyond 1 meter, limiting their ability to simulate natural focus cues. A few older HMDs, such as the Magic Leap One~\cite{MagicLeap2022}, incorporated two focal planes. Some recent video see-through HMDs, such as the Varjo XR-4 Focal\footnote{https://varjo.com/products/xr-4}, use gaze-driven auto-focus cameras to enrich the video see-through experience. However, the focal point of the display lenses is still set to a constant 140 cm, which is \additional{outside the user's arm reach, i.e., }beyond the effective range for virtual hand interaction. 

Early varifocal display systems, such as the 3D Display with accommodative compensation, addressed visual fatigue by dynamically adjusting focus~\cite{shiwa1996proposal}. Liu et al.~\cite{liuNovelPrototypeOptical2010} later introduced a liquid lens-based varifocal system, though it suffered from flicker. 
Dunn et al.~\cite{dunn2017wide} developed varifocal displays with deformable membrane mirrors, offering a wide FOV and fast depth switching, but at the cost of bulkiness.
More recent work includes Chan et al.’s~\cite{changMultifocalDisplaysDense2018} multifocal display generating 1600 focal planes per second, and Kaneko et al.~\cite{kanekoFocusAwareRetinalProjectionbased2021} with a focus-aware retinal projection display, simulating natural depth. 

Despite significant advances, challenges remain in reducing the size and complexity of these systems for consumer use. Compact approaches, such as Rathinavel et al.’s~\cite{rathinavelVarifocalOcclusionCapableOptical2019} focus-tunable optics and Zhou et al.’s~\cite{zhouDesignVarifocalMultifocal2022} liquid lens display, show promise, though precise calibration remains a challenge. While multifocal displays present multiple discrete focal planes, they may suffer from latency, visible focal transitions, and system complexity. In contrast, varifocal displays offer continuous focus adjustment, enabling smoother accommodation changes. This makes varifocal displays more suitable for dynamic interaction tasks, such as object selection, where switching to specific (yet not predetermined) depths is required.

Beyond varifocal and multifocal displays, alternative technologies such as holographic and light-field displays also offer potential solutions to the VAC~\cite{itohIndistinguishableAugmentedReality2021}. Holographic displays reconstruct wavefronts to simulate realistic depth cues, while light-field displays project multiple views to each eye, creating more natural accommodation without requiring dynamic focal adjustment. However, these technologies face challenges related to resolution, cost, and computational complexity, limiting their widespread adoption~\cite{ebner2024gaze, chakravarthula2019wirtinger}.

\subsection{Depth Perception in Stereo Displays}
\additional{Depth perception refers to the ability to perceive the three-dimensional structure of the environment, including both the relative depth between objects and their egocentric distance from the observer. It relies on multiple depth cues, including retinal cues (e.g., pictorial information and binocular disparity) and extraretinal cues (e.g., vergence and accommodation)\cite{Cutting1995a, Brown1988c, Durgin1995c, Patterson1992c, Renner2013c}. In stereo displays, extraretinal depth cues such as vergence and accommodation may become inconsistent due to the fixed focal distance of the display, resulting in the VAC, and may also be affected by visible depth quantization and the limited visual acuity of the display \cite{Kenyon2014a}.}

Our study focuses on the VAC, which arises from the fixed focal distance of stereo display systems in AR HMDs. Issues caused by the VAC in the human ocular system include: 1) depth perception difficulties~\cite{Durgin1995c}, 2) eyes converging closer than necessary~\cite{Iskander2019a}, and 3) visual fatigue due to discrepancies between focal and vergence distances~\cite{Hoffman2008a}. These issues degrade visual performance~\cite{Vienne2014a} and increase cognitive load~\cite{Daniel2019}.
For example, Erkelens et al.~\cite{erkelens192VergenceAccommodationConflicts2020} used a varifocal display to define the zone of clear vision, i.e., the range where the VAC is tolerable before perceived image quality changes. This zone encompasses up to 0.5 diopters on both sides of the focal plane. Their setup used an additive multiplane haploscope to test three vergence distances with positive, neutral, and negative displacement from a fixed focal distance.
Similarly, Iskander et al.~\cite{Iskander2019a} investigated the impact of Virtual Reality (VR) on eye vergence using biomechanical simulations with consumer-grade HMDs. Participants focused on a cube at varying distances, and eye-tracking was used to compare vergence angles under conflicting (VR) and ideal (natural) viewing conditions. Results showed greater variability in vergence movements in VR, leading to incorrect depth perception and visual fatigue. Increased object depth further reduced eye-gaze performance. Their approach avoided complex setups but required compensating for head movement, adding analytical complexity.
However, all these studies focus on depth perception issues and do not address interaction. Our work extends this line of research by analyzing interaction performance in a varifocal AR setup.

\subsection{3D Pointing}
Our study focuses on 3D pointing, where users point to a target in space before selecting it, typically by pressing a button or making a gesture. This task is both a visual and biomechanical challenge, as users must identify the target's position in 3D space and then move their arms to reach it. 

In stereo displays, the presence of the VAC negatively affects user performance in 3D selection tasks, particularly when pointing at targets at varying visual depths. Previous studies have shown that with the virtual hand technique, \additional{i.e., direct target selection within arm’s reach,} movement time (MT) and throughput (THP) are lower for movements in visual depth compared to lateral movements on a large stereo display \cite{barreramachucaEffectStereoDisplay2019a}. 
Their comparison with a real-world setup, which yielded the opposite results for the same tasks, indicated that the limitations of stereoscopic displays might be the underlying reason.
Batmaz et al.~\cite{batmaz2019:DoStereoDeficinciesAffesct} verified the presence of this effect in modern AR and VR HMDs. For raycasting, previous studies indicated that varying target depth affects performance~\cite{Teather:2011:Pointingat3Dtargets} and that this performance loss is a consequence of the VAC~\cite{batmaz2023virtualhandvac, Batmaz:2022:VRST}. Other research has identified that movement biomechanics also affect 3D pointing performance~\cite{Batmaz:2022:VRST, 10.1145/3544548.3581191}. However, all of these studies utilized commercial fixed-focal stereo displays. 

Finally, Batmaz et al.~\cite{Batmaz:2022:VAC} uniquely utilized a custom-made multifocal display apparatus that eliminates the VAC. They found that using a multifocal display reduces the effect of the VAC on 3D selection with the \emph{virtual hand} technique. McAnally et al.~\cite{mcanally2024visually} also examined the effect of different accommodative-demand conditions on interaction performance, but used fixed display focal distances combined with optical correction rather than dynamic focal adjustment. Aside from these studies, we found no other work that explored the effect of the VAC on user interaction using displays that actively and dynamically adjust focal depth during interaction. Our work adds novel research in this gap by conducting a user study on 3D pointing performance using a varifocal AR display that dynamically switches focal depth.

\section{Varifocal Display Design}
We developed a varifocal stereo display to enable us to perform a user study based on the ISO 9241-411~\cite{ISO2015} methodology to investigate the effect of the VAC on 3D pointing. This section details our design decisions and display characteristics.

\begin{figure}[!t]
\centering
\includegraphics[width=\linewidth]{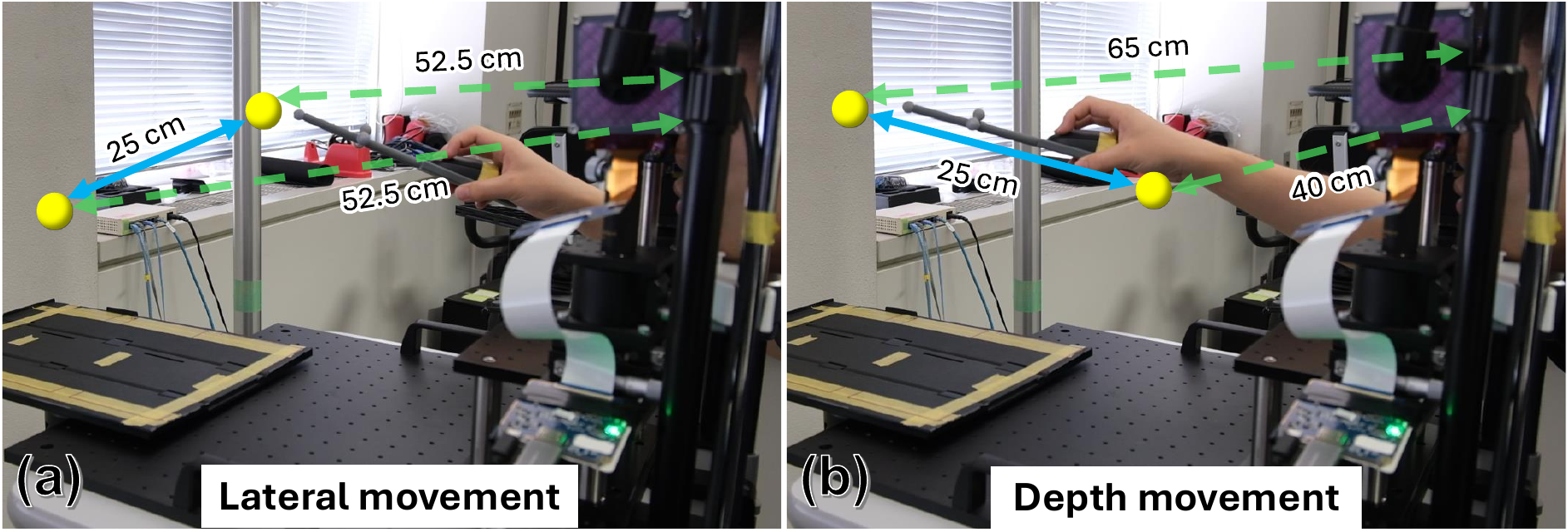}
\caption{Movement directions. (a) Lateral direction, with both virtual targets positioned 52.5 cm from the user; (b) depth direction, with virtual targets positioned 40 cm and 65 cm from the user, respectively. For both movements, the separation distance between the virtual targets is 25 cm.}
\label{fig:movement}
\end{figure}

\subsection{Design Considerations} \label{sec:systemdesign}
Our study directly replicates Batmaz et al.'s \cite{Batmaz:2022:VAC} multifocal experimental setup and 3D selection task within a varifocal display system, allowing us to compare our results directly to their work. This approach also allows us to compare with results from Barrera Machuca and Stuerzlinger~\cite{barreramachucaEffectStereoDisplay2019a} and Batmaz et al.~\cite{batmaz2019:DoStereoDeficinciesAffesct}, who used commercial stereo displays. All these studies employed a variant of the ISO 9241-411 selection task~\cite{ISO2015}, where pairs of targets were positioned along a single movement direction, either laterally or along the line of sight, i.e., in (visual) depth (\autoref{fig:movement}).
\additional{Pilot testing showed that some participants could not comfortably reach the farthest targets. Therefore, we chose the target distances of 40, 52.5, and 65 cm to better match participants’ reachable workspace.}
Thus, and also considering the sizes of the targets (1.5, 2.5, and 3.5 cm), the apparatus must support a minimum horizontal FOV of \(30.4^\circ\). In addition, presenting virtual targets with depth cues requires that the display provide stereoscopic views with a sufficiently large FOV, which is typically considerably smaller than the monocular FOV of the display within the interactive range. 
\additional{Based on geometry,} for a user with an IPD of \(63\) mm, an AR display that supports a stereo FOV of \(30.4^\circ\) at \(52.5\) cm distance thus requires a monocular FOV of \(36.7^\circ\).

Another important consideration is the replicability of the system. To ensure replicability, the apparatus was constructed using commercially available optical components.
Liu et al.'s design~\cite{liuNovelPrototypeOptical2010}, which our implementation is based on, is an optical see-through configuration that uses a beam-splitter to combine real and virtual imagery, resulting in a stereo AR display.
Their design supports a wide FOV, allowing virtual targets to appear with sufficient lateral separation, and has low complexity, facilitating calibration with only basic optics knowledge. Their design also employs a liquid-crystal tunable lens, a component now available in commercial versions with switching times around 2.5~ms, making it suitable for interactive studies involving depth changes. We further adapted this optical configuration for AR scenarios and for studying interaction performance, enabling consistent evaluation of VAC-related effects.

Other considerations include image clarity, where we focus on offering a high display resolution to support (a) clear stereo vision, (b) reducing the likelihood of diplopia by keeping targets at a safe viewing distance that supports comfortable binocular fusion, and (c) avoiding interocular crosstalk, where the image for one eye is seen by the other one~\cite{Patterson2009}. Moreover, we intentionally removed image-based depth cues, such as shadows and texture, to eliminate confounding factors. Finally, our display is fixed to a table to avoid the challenges associated with HMDs, such as weight issues that can affect how a display sits on a user’s head~\cite{Buck2018}. We also use a chin and forehead rest to eliminate participants’ head movements, which might affect the perception of the stereo display.

\begin{figure*}[!t]
\centering
\includegraphics[width=\linewidth]{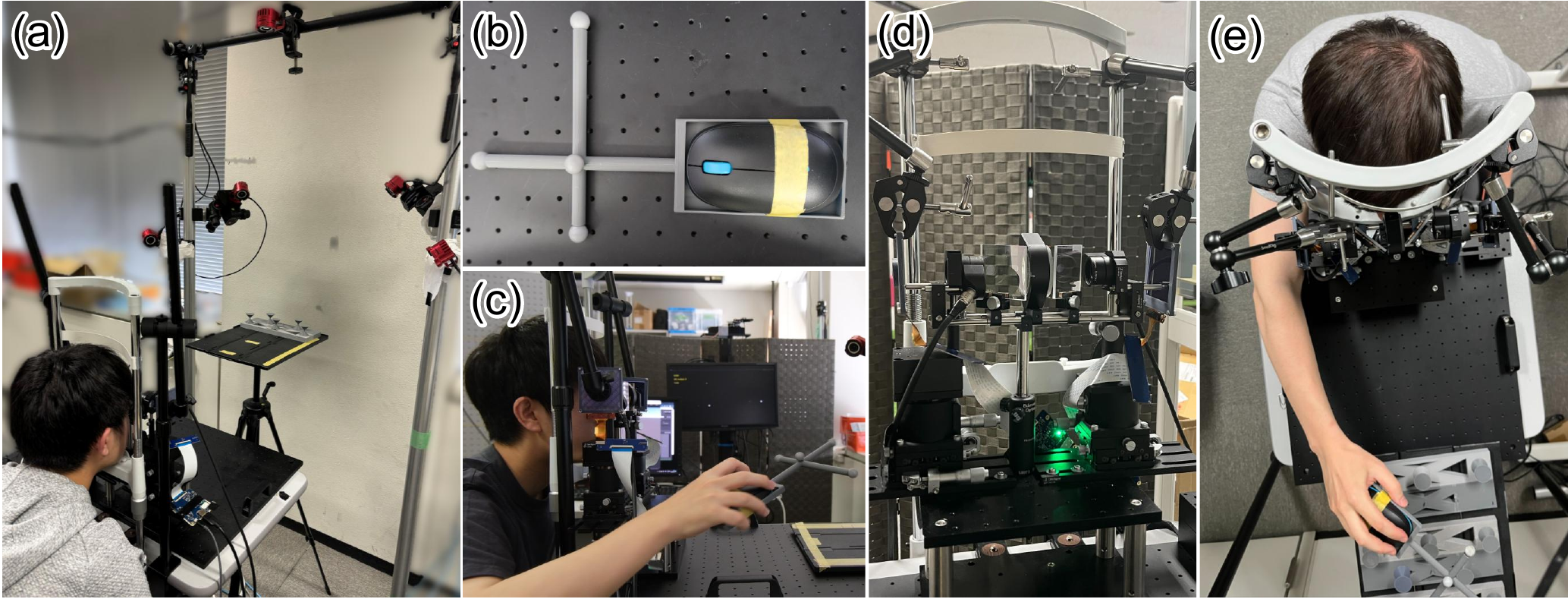}
\caption{User study. (a) Experimental environment. (b) Wand. (c) An image of a user performing our experiment. (d) The revised setup, which improves mechanical robustness and ergonomics. \additional{(e) A user performing the experiment while extending their arm within the revised setup, which offered approximately 10~cm more interaction space.}}
\label{fig:user}
\end{figure*}

\subsection{Implementation}

\begin{figure}[!t]
\centering
\includegraphics[width=\linewidth]{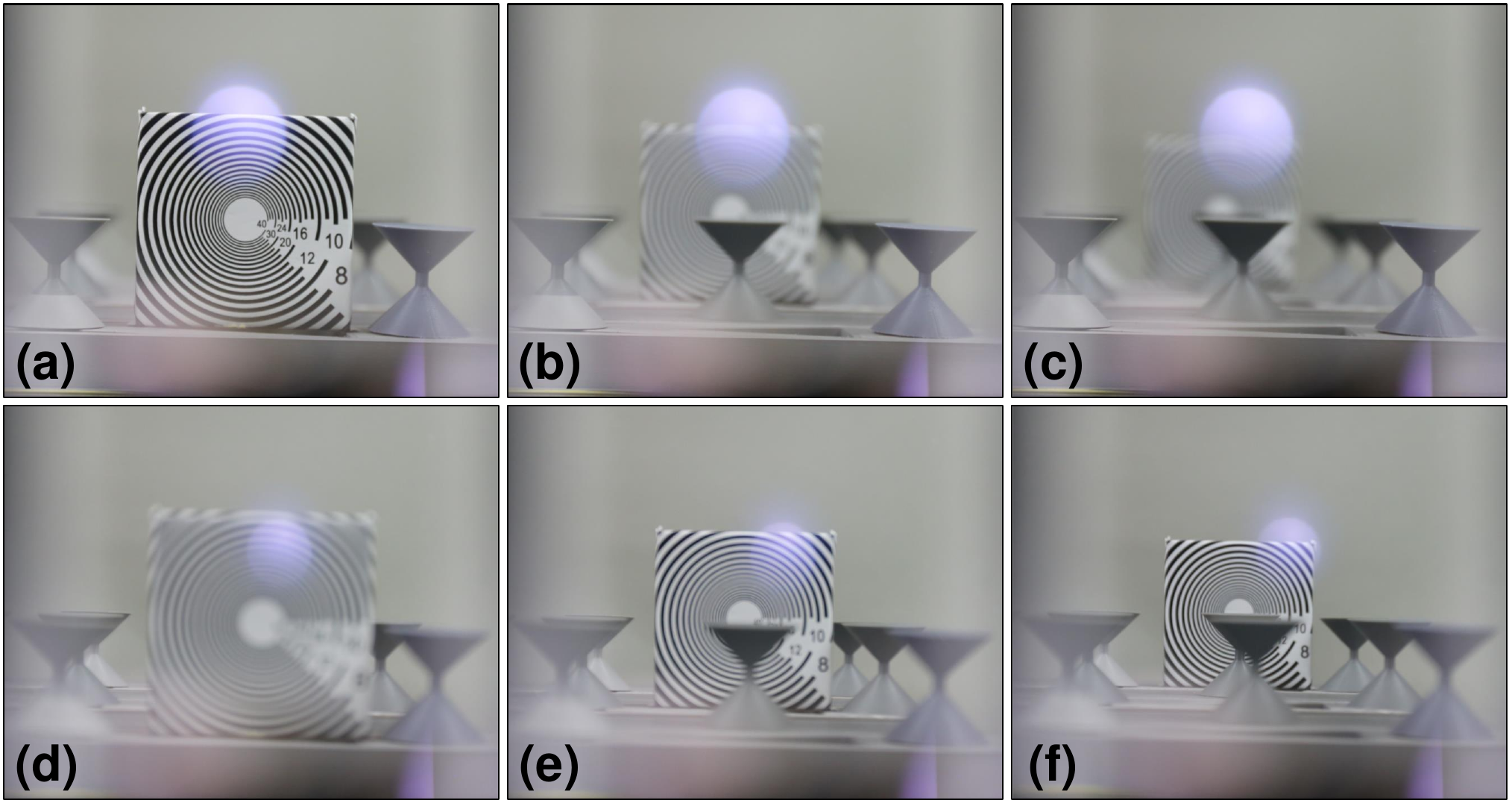}
\caption{Exemplary views with varifocal feature activated. The user focuses on the target shown at different distances: (a-c) at the closest plane, with the test pattern at the (a) closest, (b) middle, and (c) farthest plane; (d-f) at the farthest plane, with the test pattern at the (d) closest, (e) middle, and (f) farthest plane.
}
\label{fig:targets}
\end{figure}

To satisfy the task requirements, we implemented our apparatus accordingly while also following the varifocal stereo display design described above. The same optical configuration and varifocal principle were used across both user study iterations, ensuring that any observed differences are not attributable to changes in optical behavior.

\subsubsection{Display}
\autoref{fig:teaser} (right) shows our binocular benchtop prototype built using off-the-shelf components. We used two Optotune EL-10-30-Ci-VIS-LD focus-tunable lenses, each with a 10 mm clear aperture and an optical power of +5D to +10D. With a 2.5 ms response time, these lenses ensure fast accommodation for interaction. Each lens holder contains a 25 mm diameter, 25 mm focal length aspheric objective lens. Through the use of 50 mm spherical mirrors (radius of curvature: 60 mm, focal length: 30 mm), the system achieves an accommodation range of 0D to 5D, equivalent to focusing from infinity to 20 cm.

We placed these two spherical mirrors back-to-back in an 18 mm-thick lens holder. To minimize the distance between the eyeboxes (the region where the user's eyes can perceive the full image) and match the average human IPD (63 mm)~\cite{dodgson2004variation}, the lens holder was thinned to reduce the combined mirror thickness. The mirrors were secured using double-sided tape.

Two plate beam splitters (\(50 \times 50\) mm) were placed in square optics holders at 45$^{\circ}$ and 135$^{\circ}$ angles to the main optical axis. Fixed with two cage guide rods and spherical mirrors, this setup offers adjustable eyebox spacing by translating the beam splitters and holders along the optical axis. Additional spacing adjustments are possible by shifting the beam splitters within the holders. Our current design supports an IPD range of 63–70~mm.

As image sources, we used two Sharp LS029B3SX02 LCDs with 1440×1440 resolution, a 51.84 × 51.84~mm active area, and a 120~Hz refresh rate.

To accommodate each user's upper body size, the entire system was fixed on a height-adjustable table, with a chin rest clamped to the edge. 
To reduce fatigue, which could affect results, we also used a height-adjustable chair to ensure ergonomic comfort.
Across the two study iterations, the optical configuration and varifocal principle remained unchanged. In the initial setup (\autoref{fig:user}(a)), the two LCDs were secured using camera clamp arms mounted on the table, whose vertical posts partially constrained lateral arm movements. In the revised setup (\autoref{fig:user}(d)), the LCDs were instead suspended using articulated mounting arms attached to the chin-rest assembly, increasing the unobstructed interaction space toward the user by approximately 10~cm (\autoref{fig:teaser}(left), \autoref{fig:user}(a)). In addition, the cage system was reinforced with additional cage plates and rods to ensure stable lens alignment under incidental contact. These changes improved mechanical robustness and ergonomics but did not alter the display's optical behavior, focal planes, or rendering pipeline.
While our current implementation is benchtop-mounted to ensure optical and mechanical stability for the user study, the optical layout is compatible with wearable implementations, as shown in prior work~\cite{ebner2022video}.

\subsubsection{Tracking System}



We used five OptiTrack Flex13\footnote{https://optitrack.com/cameras/flex-13} cameras (120 FPS, 1280×1024 resolution, 8.3 ms latency) with Motive\footnote{https://optitrack.com/software} 1.9.0 for motion capture.

The cameras were arranged as shown in \autoref{fig:user}(a): two capturing side views, two capturing front views, and one overhead, mounted on a crossbar. The four side and front cameras were attached to poles extending from the ceiling to the floor.

For the pointing device (\autoref{fig:user}(b)), we designed a $25\times15$~cm wand, weighing 32~g, with four reflective balls for optical tracking. A groove at the bottom held a small mouse, bringing the total weight to 108~g. To help users see their 3D selection position without VAC-related optical conflicts, a 5~mm cursor was displayed 5~cm in front of the top reflective ball. Users held the mouse as shown in \autoref{fig:user}(c) and selected virtual targets by clicking the left button. The wand was tracked in six degrees of freedom, allowing participants to move their hand freely in 3D space during the task.

\subsubsection{Software}
We implemented target and cursor rendering, along with tunable lens power adjustment for target locations, in Unity 2019.2.10. To render the views for both eyes at three focal planes, six virtual cameras were positioned relative to the user's head. These cameras were calibrated using off-axis projections to align virtual and real images. As described in Section~\ref{sec:systemdesign}, four virtual targets were placed at 40, 52.5, and 65 cm from the cameras. At 40 and 65 cm, one target was centered, while at 52.5 cm, two targets were positioned 25 cm apart on either side of the central axis (\autoref{fig:movement}). Each target had three diameters: 1.5, 2.5, and 3.5~cm. \additional{The cursor was rendered as a white sphere with a diameter of 0.5~cm and remained visible throughout the task. Because the targets were opaque, the cursor was naturally occluded when it moved inside a target. To reduce the likelihood of conflicting visual cues in some conditions, the cursor was positioned approximately 2~cm above the tip of the wand.}

In the experiment, only one target was visible at a time, but targets alternated between two fixed positions along the movement direction, and users were instructed to select it as quickly and accurately as possible. 
If a target was missed, an error sound played through the PC’s speakers. For details on the procedure, see Section~\ref{sec:procedure}.
Based on prior work on pointing movements~\cite{batmaz2019:DoStereoDeficinciesAffesct, Batmaz:2022:VAC}, \additional{gaze typically precedes the hand and is directed toward the target location during such movements~\cite{tramper2011visuomotor}. Therefore,} we assume users \additional{primarily} focus on each newly appeared target rather than the cursor. This aligns with Fitts' law, which emphasizes rapid, aimed movements toward a target to hit it. Consequently, we did not include eye-tracking, avoiding potential tracker inaccuracies~\cite{dunn2019required} and tunable lens delays. This also simplifies lens control, as the optical power only needs to match the depth of the newly appeared target. Under the assumption that participants verged on the new target, adjusting display focus to the new target would have reduced the VAC. \additional{Accordingly, the present implementation should be interpreted as a target-depth-driven varifocal control strategy rather than a fully gaze-contingent varifocal system.}

We used microcontrollers with a virtual COM port driver to control the lens drivers and achieve the desired optical power. We adjust the optical power of the tunable lens by varying the voltage applied to the liquid crystal, altering its thickness. However, even identical lenses may vary due to differences in production, which can be amplified by refraction and reflection in our system. To address this, we performed secondary calibration using manufacturer-provided parameters. The tunable lenses were connected to a camera with a SONY IMX250 sensor (2448 × 2048 pixels) and a 35 mm/F1.65 lens, observing a resolution test pattern to determine the voltage value that provided the sharpest image at the desired target depths. These values were then used to set the voltage during the experiment. In varifocal mode, the voltage was dynamically adjusted to match the user’s focus at 40, 52.5, and 65 cm, while in fixed-focal mode, it remained fixed at 52.5 cm. Exemplary views with the varifocal feature at different planes are shown in~\autoref{fig:targets}.

The experiment ran on a PC with an Intel Core i5-13400F processor, 16GB of RAM, and an NVIDIA GeForce RTX 4070 GPU, running on Windows 11.

\begin{figure}[!t]
\centering
\includegraphics[width=\linewidth]{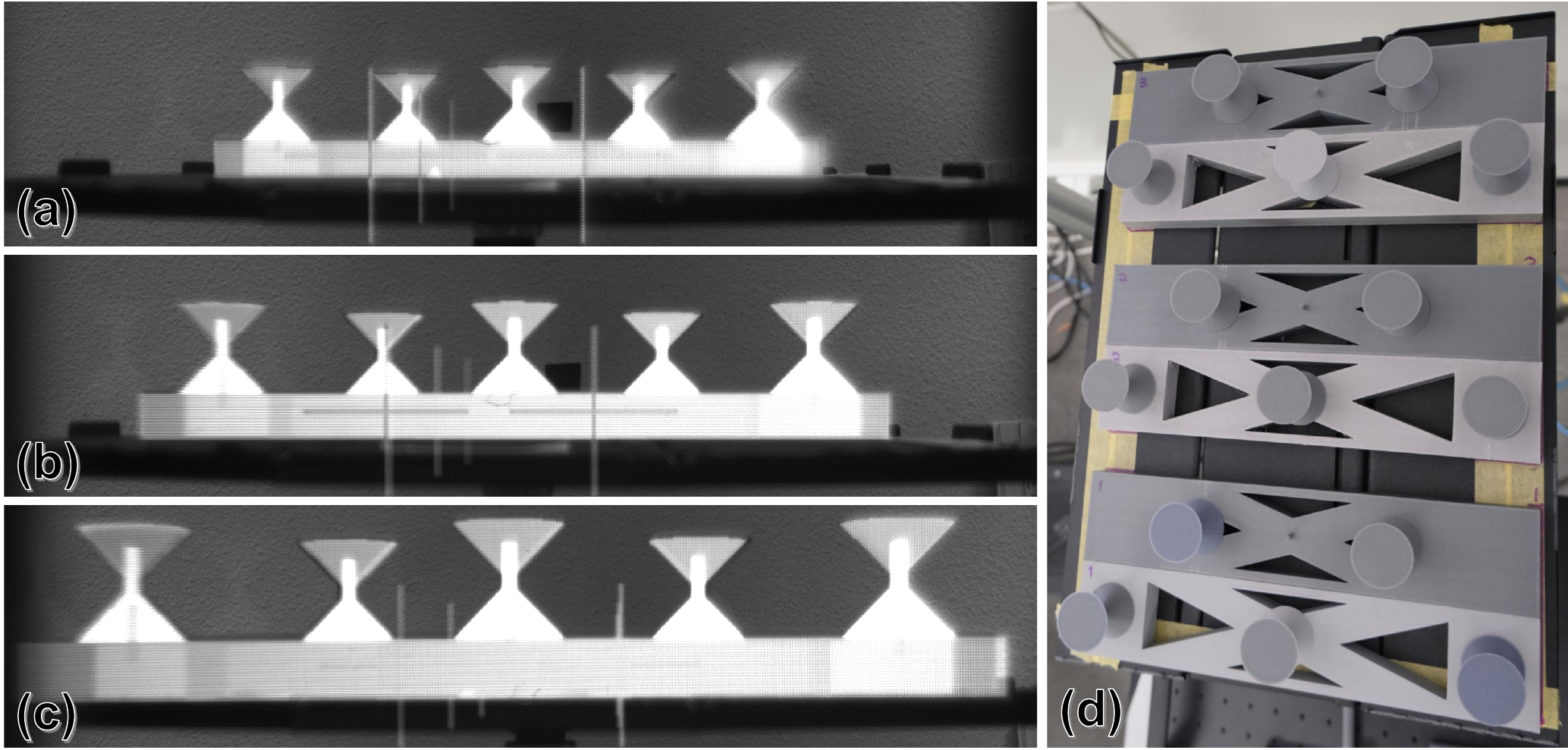}
\caption{Calibration views from the user's perspective, with the calibration platforms positioned at (a) 65 cm, (b) 52.5 cm, and (c) 40 cm from the user. The virtual and real calibration platforms are aligned at these points. (d) The three calibration platforms were placed on a height-adjustable table, allowing easy removal after each depth was calibrated.}
\label{fig:calibration}
\end{figure}

\subsection{Calibration}
Since the experiment requires the user to select virtual objects, we need to ensure that virtual objects are perceived at depths that correspond as closely as possible to equivalent real objects.

\subsubsection{Calibration Platforms}
Based on the approach pioneered by Batmaz et al.~\cite{Batmaz:2022:VAC}, we used a 3D printer (AnkerMake M5\footnote{https://www.ankermake.com/m5}) to create three calibration platforms (\autoref{fig:calibration}) and hourglass-shaped calibration targets for aligning the virtual and real views. These platforms were placed on a height-adjustable table at distances of 40, 52.5, and 65 cm from the eyebox, corresponding to the depths of the virtual targets in the experiment.

The calibration platform consists of two layers, front and back. The lateral length of the platform is 25~cm, matching the range where the virtual targets appear, and its longitudinal depth is 10~cm, with each layer being 5~cm deep. This allows for quick verification: if the user’s virtual view does not fully cover the platform, the FOV is insufficient. Three hourglass-shaped objects (3~cm in diameter, 5~cm in height) were arranged on the front layer, with two identical ones filling the gaps on the back layer. The benefit of the hourglass shape is that it supports accurate visual judgments from both viewpoints, making it easier to see misalignments along the depth dimension~\cite{Batmaz:2022:VAC}.

The use of a height-adjustable table allows us to lower the setup after calibration so users can perform the pointing task directly, maintaining platform stability without interfering with pointing actions.

\subsubsection{Compensation for Transversal Magnification}
As the optical power of the tunable lens changes during accommodation, the transversal magnification of the virtual view also varies~\cite{liuNovelPrototypeOptical2010}. We thus adjust the size of the views according to the transversal magnification at each depth to ensure that the virtual targets and cursor appear at the correct place and with the intended size to the user at different depths.
According to the unfolded optical path diagram in Figure 12 of Liu et al.'s paper~\cite{liu2008optical}, we can derive the transversal magnification $M$ as: 
\begin{equation}
    M = \frac{u'}{u} \cdot \frac{v'}{v},
\end{equation}
with parameter definitions consistent with the original paper: \( u \) is the distance from the LCD to the lens group (including the objective lens and the tunable lens), and \( u' \) is the image distance formed by the lens group, i.e., the distance from the image \( I' \) to the lens group. \( v \) is the distance from \( I' \) to the spherical mirror, while \( v' \) is the image distance formed after reflection by the spherical mirror, i.e., the distance from \( I'' \) to the spherical mirror. 
As we focus on magnification variation, directional signs are omitted.

Since our goal is to ensure that the size of the virtual content appears consistent, we only need to determine the relative magnification between target distances. Taking the close and middle target distances as an example, the ratio of their transversal magnifications, \( M_c \) and \( M_m \), is given by:
\begin{equation}
    \frac{M_c}{M_m} = \frac{ (d_c - R)^2 \cdot d_m } {(d_m - R)^2 \cdot d_c},
    \label{eq:relaMag}
\end{equation}
where \(d_c = 40\) cm, \(d_m = 52.5\) cm, and \(R\) is the radius of curvature of the spherical mirror. 
Thus, during calibration, we only need to align the virtual view with the real calibration platform at a single target distance as a reference. Then, by applying the calculated relative transversal magnifications, we can compensate for the virtual views at the remaining two target distances, ensuring greater calibration precision.

\section{User Studies}
The main goal of our user studies was to compare user performance in virtual hand pointing at targets at different depths with and without the varifocal stereo display.

\subsection{Overview}
We conducted two iterations of the same user study protocol.
Study 1 used the initial benchtop setup, while Study 2 employed an ergonomically revised and mechanically reinforced apparatus.
Both studies followed an identical task design, stimuli, and procedure. As past work had found that the movement biomechanics affect user performance~\cite{10316355}, study 2 served as a validation under improved ergonomic conditions, \additional{rather than as a fully powered confirmatory replication.} This helped verify that the trends observed in Study 1 were not solely attributable to the constraints of the original benchtop apparatus.

\subsection{Hypotheses}
\textit{H1 - Varifocal stereo display positively affects user performance for 3D target selection at different visual depths in peri-personal space compared to fixed-focal stereo displays.} Here, ``positively affects" refers to the performance metrics detailed in Section~\ref{sec:measurements}, including movement time, throughput, accuracy, and fatigue-related effects. We expect that the varifocal stereo display mitigates the VAC, which has been shown to impair depth perception and increase visual fatigue. More importantly for our work, prior studies using fixed-focal stereo displays have consistently reported degraded interaction performance due to the VAC~\cite{batmaz2019:DoStereoDeficinciesAffesct, 10.1145/3611659.3615686, barreramachucaEffectStereoDisplay2019a, batmaz2023virtualhandvac, Batmaz:2022:VRST}. Our system is designed to reduce this mismatch, and the hypothesis aims to quantify how varifocal displays overcome these limitations during virtual hand selection.

\textit{H2 - Using a varifocal stereo display affects user hand movements when selecting 3D targets in peri-personal space compared to fixed-focal stereo displays.} As varifocal displays can reduce the accommodative mismatch, we hypothesize that arm movements during interaction do not present the typical second sub-movement found by Barrera Machuca and Stuerzlinger in fixed-focal stereo display systems~\cite{barreramachucaEffectStereoDisplay2019a}. That second sub-movement indicates a depth-estimation error that extends the correction phase of the movement. The importance of \textit{H2} is that it investigates the dissimilarity between arm movements when using a varifocal display and a fixed-focal one. 

\subsection{Methodology}
We study 3D target selection via a Fitts’ law task, which first involves using a wand to move the cursor into the target and then using a button to select the target. The experiment was reviewed and approved by the Institutional Review Board for studies with human participants from the Nara Institute of Science and Technology.

\subsubsection{Participants}
\paragraph{Study 1}
We recruited 24 participants from the local university community (17 male, seven female) aged 22-38 years (M = 27.04, SD = 4.66). All participants used their dominant hand to do the task (with two being left-handed). All the participants had normal or corrected-to-normal vision. However, due to the limited FOV supported by our prototype, we had to shorten the eye relief (the distance between the user's eye and the surface of the holder of the spherical mirror) to achieve the maximum possible FOV. Such a short eye relief usually cannot accommodate the nose bridge of eyeglass frames. However, in the experiment, we found that if a user's IPD is greater than 68~mm, our system offers sufficient FOV with this longer eye relief, even for people wearing eyeglasses. Therefore, two nearsighted participants were able to wear their regular glasses, while the others wore contact lenses to correct their vision.

Participants had varying familiarity with 3D content interaction. Regarding 3D games, 16 participants did not play 3D games; five participants played 3D games for 1-5 hours per week, two participants played for 6-10 hours per week, and one participant played for more than 10 hours per week. Additionally, six participants used 3D CAD systems for 1-5 hours per week, and one participant used them for 6-10 hours per week. Regarding experience with immersive systems, 20 participants had experienced VR at least once, including 14 who had also experienced AR. One participant had experienced only AR at least once.

\paragraph{Study 2}
We recruited an additional 12 participants (seven male, five female) from the same population as Study~1, aged 23--31 years (M = 25.3, SD = 2.73). All participants used their dominant hand for the task (two were left-handed) and had normal or corrected-to-normal vision. The same inclusion criteria and vision correction requirements as in Study~1 were applied. Participants reported varying prior experience with 3D content: four reported no 3D game play, six reported 1--5 hours/week, and two reported 6--10 hours/week; nine reported no 3D CAD use, while three reported some CAD experience (1--3 times/week, 1--5 hours/week, or 6--10 hours/week; one participant each). Five participants had used a VR HMD at least once (three 6+ times, two 1--3 times), and three had used an AR HMD at least once (two 6+ times, one 1--3 times).

\begin{figure}[!t]
\centering
\includegraphics[width=\linewidth]{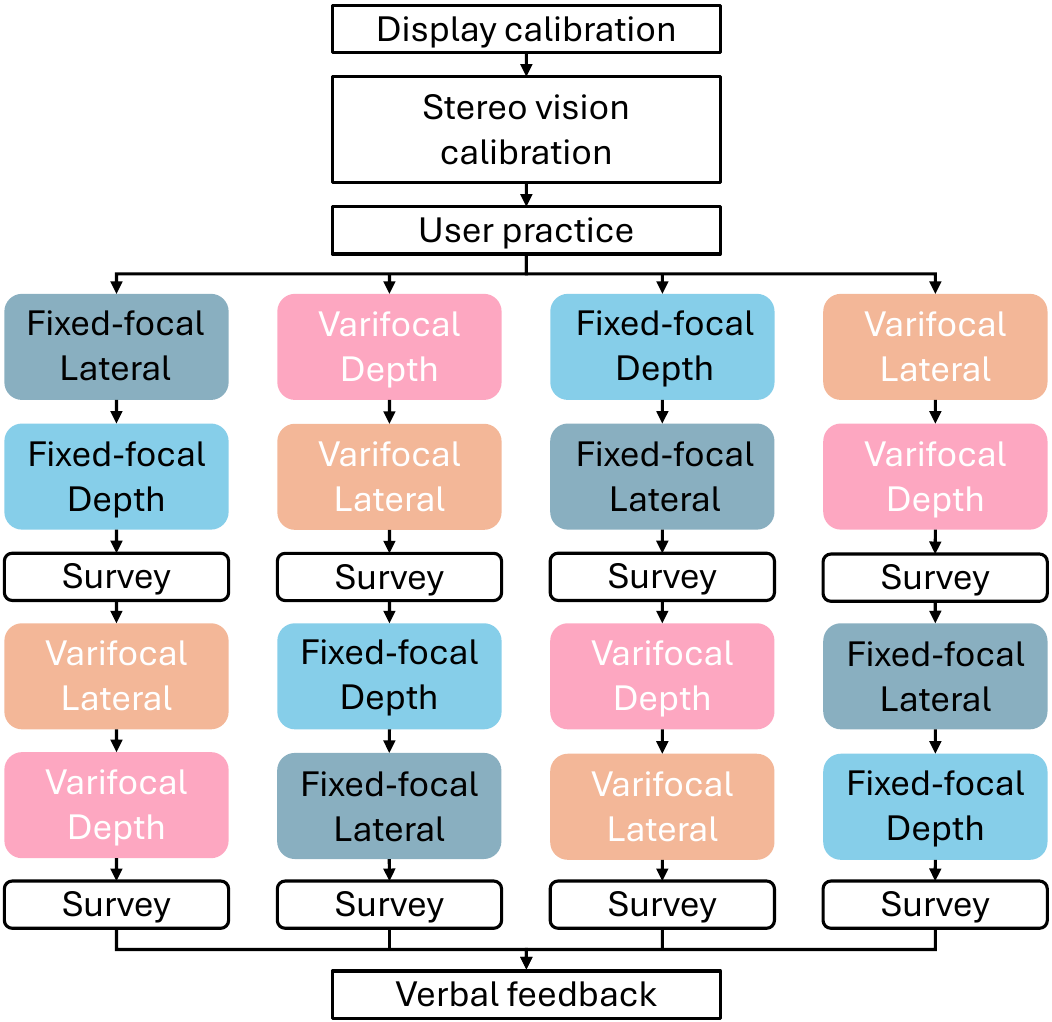}
\caption{Experimental procedure. After the practice step, the task sequence for each user was counterbalanced with a Latin square order.
}
\label{fig:flow}
\end{figure}


\subsubsection{Procedure} \label{sec:procedure}
After entering the lab, participants received an explanation of the display calibration and experimental task. Spherical targets were positioned along a single movement direction for the experimental task. In the \textit{lateral movement}, targets appeared alternately, requiring users to move the wand between the left and right, or in the \textit{depth movement} between the front and back targets (\autoref{fig:movement}). The start position in each trial was naturally determined by the endpoint of the previous motion.

Once consent was obtained, participants' IPD was measured, they answered a demographics questionnaire, and then we proceeded to calibrate the display to ensure correct optical stimuli for each participant. During the calibration, the participant's head was first positioned so that the images of the displays were visible to both eyes. This was achieved with the help of a chinrest and by adjusting the beamsplitters. Afterward, the three physical 3D-printed calibration platforms (\autoref{fig:calibration}(d)) were used to calibrate the depth and spatial distribution of the displayed objects, as explained above. During the calibration task, the system operated in the fixed-focal mode, with the tunable lens focus fixed at the middle depth of 52.5 cm. The calibration task involved aligning the physical object and its virtual image (\autoref{fig:calibration}(a-c)). The calibration was done for one eye at a time to account for physical differences in the display lenses. After the calibration was finished, participants could practice both types of pointing tasks for as long as needed to get used to them. Our instruction to participants emphasized giving speed priority while still maintaining accuracy. To reduce fatigue, participants were required to take a mandatory break of 15 seconds between different movement types and 45 seconds between different display conditions, where we showed a screen titled “REST”, and a rest between conditions for as long as needed. Additionally, participants were permitted to take a two-minute break between trial rounds whenever they felt tired.

After each condition of the experiment, participants filled out a survey that asked about the ease of use, perceived speed, and fatigue level. \additional{The interval was typically approximately 2-3 minutes, depending on the time taken to complete the questionnaire and the optional rest period.} At the end of the experiment, we also asked them about their preferred condition. Including the calibration time, the experiment took approximately 60 minutes. The overall experimental procedure is illustrated in~\autoref{fig:flow}.

\subsubsection{Measurements} \label{sec:measurements}
Following previous work by Batmaz et al.~\cite{Batmaz:2022:VAC} and Barrera Machuca and Stuerzlinger~\cite{barreramachucaEffectStereoDisplay2019a}, we recorded and analyzed the following measures: 

\paragraph{Target Selection Performance} We used three different measurements to identify user selection performance. First, we measured the \textit{MT (s)}, which is the time it takes a person to move their arm to select a target. Then, we recorded the \textit{Error rate (\%)}, which is the number of incorrect selections in a set of targets. Finally, we calculated the \textit{THP (bps)} using the Shannon Formulation of Fitts' Law, based on previous work~\cite{teather2009effects, Teather:2011:Pointingat3Dtargets}. THP combines speed and accuracy, making this performance metric less dependent on user strategies. Fitts' Law~\cite{fitts1954information} predicts the 3D pointing MT, i.e., the time between the start of a movement and the (successful) selection of a target \additional{via button press}. See Equation \autoref{mtOne} for the Shannon Formulation~\cite{mackenzie1992fitts}:

    \begin{equation} \label{mtOne}
        \mathit{MT} = a+b\cdot\log_2 \left( \frac{D}{W}+1 \right)  = a+b\cdot \mathit{ID}
    \end{equation}

In Equation \autoref{mtOne}, the logarithmic term, known as the index of difficulty ($\mathit{ID}$), indicates the overall pointing task difficulty. The $\mathit{ID}$ is calculated from D and W, which are the target distance and size, respectively, while a and b are empirically derived via linear regression. We then calculated THP based on effective measures as defined in the ISO 9241-411:2015 document~\cite{ISO2015} (\autoref{thpOne}):

    \begin{equation} \label{thpOne}
        \mathit{THP} = \frac{\mathit{Effective Index Of Difficulty}}{\mathit{Movement Time}} = \frac{\mathit{ID}_{e}}{MT}
    \end{equation}

Equation \autoref{ideOne} below defines the effective index of difficulty ($\mathit{ID}_e$), where $A_e$ is the effective movement amplitude (the distance from the start of the selection movement to the selected position) and $W_e$ is the effective target width (the standard deviation between the selected position and the target center, $\mathit{SD}_x$) and characterizes the accuracy of the task performance~\cite{MacKenzie:2008:SpeedAccuracyTradeoff, mackenzie1998comparison}: 

    \begin{equation} \label{ideOne}
        \mathit{ID}_e = \log_2 \left( \frac{A_e}{W_e} + 1 \right) = \log_2(\frac{A_e} {(4.133 \cdot \mathit{SD}_x)} +1)
    \end{equation}

\paragraph{Movement Paths} We analyzed the movement paths using five measures. \textit{Target re-entry} is the number of times the cursor sphere enters the target in each click, while \textit{Speed} is how fast the user moves their arm to reach the target. \textit{Ballistic time} is the time it takes to do the initial impulse of a rapid aimed movement, \textit{Correction time} is the time of the guided final control movement, and \textit{Correction distance} is the distance of this guided final control movement. We calculated these measures based on Nieuwenhuizen's work, \additional{in which movement trajectories were segmented based on the velocity profile. The ballistic phase was defined as the initial impulse up to peak velocity, and the correction phase as the subsequent visually guided adjustment phase until target selection~\cite{Nieuwenhuizen2009c}.}

\paragraph{Self-reported Preferences} We asked our participants about the ease of interaction and perceived speed in each experimental condition (varifocal versus fixed-focal display and depth movements versus lateral movements). We also asked participants about their perceived motion sickness and fatigue in each condition. Finally, we asked about their preferences and whether they could identify the display type they used during the studies. Participants' answers were collected using a custom version of the Lime survey.

\subsection{Experimental design}
We performed a two-factor within-subjects user study with two different \textbf{display} conditions ($2_\mathit{DISPLAY}=$ Varifocal and Fixed-focal) and two \textbf{movement direction} ($2_\mathit{MD}=$ Lateral and Depth), which resulted in an experimental design with ($2_\mathit{DISPLAY} \times 2_\mathit{MD}$) four conditions. We counterbalanced the display and movement direction conditions with a Latin Square across participants, \additional{mitigating possible order effects between display modes, including potential short-term adaptation effects in vergence–accommodation coupling reported in previous work~\cite{ Iskander2019a, wang2024prolonged}.} To vary the task difficulty, we used 3 $\mathit{ID}$s, using all combinations of three \textbf{target sizes} ($3_\mathit{TS}$) and one \textbf{target distance} ($1_\mathit{TD}$). In total, each participant performed $2_\mathit{DISPLAY} \times 2_\mathit{MD} \times 3_{\mathit{ID}} \times $ 11 repetitions $\times$ 3 rounds = 396 trials. Targets were placed in the X-Z plane, following conventions in prior interaction studies that focus on comparisons of depth and lateral pointing~\cite{murata:2001:extendingFitts}.

\section{Analysis and Results}

\subsection{Overview}
We analyzed task performance using linear mixed-effects models (LMM) to account for the repeated-measures design and inter-participant variability. 
\additional{Time-related measures (Movement Time, Ballistic Time, and Correction Time) were log-transformed before the LMM analysis because they showed positive skewness, and the transformation improved the distribution of model residuals.} For each dependent variable, display mode, movement direction, and ID were treated as within-subject fixed effects, including all interactions. \additional{The fixed-focal condition was used as the reference level for display, the lateral condition as the reference level for movement direction, and the lowest ID level as the reference for ID.} Participant was included as a random effect with a random intercept and a random slope for display mode ($1+display~\mid~participant$), to account for baseline performance differences and individual variability in the display effect.

Statistical significance of fixed effects was evaluated using analysis of variance on the fitted mixed-effects models. To characterize display-related differences, estimated marginal means (EMMs) were computed for display while averaging over axis and ID in the model, and the varifocal--fixed contrast was used to quantify the direction and magnitude of display effects, reported as Cohen’s $d$. For each main-effect block, we additionally report semi-partial $\Delta R^2_m$ as the decrease in marginal $R^2$ after removing the corresponding fixed-effect block~\cite{nakagawa2013general, stoffel2021partr2}. An alpha level of 0.05 was used for all statistical tests.

\autoref{tab:LMM} summarizes the main effects obtained from the LMM analysis for Study~1. Strong effects of movement direction and ID were observed across most dependent variables, confirming the expected influence of task structure on performance.

\begin{table*}[!h]
\footnotesize
\centering
\caption{LMM results for Study~1 (24 participants). Only the main effects of display mode, movement direction, and ID are reported. Significant results highlighted in bold.
Effect sizes are reported as Cohen’s $d$ for the display contrast (varifocal--fixed) and semi-partial $\Delta R^2_m$ for each main-effect block.
Interaction terms are omitted for brevity and instead described in the main text in detail; their $\Delta R^2_m$ values were generally small (largest observed $\approx 0.012$).}
\label{tab:LMM}
{%
\begin{tabular}{|c|c|c|c|}
\hline
 & Display & Movement direction & ID \\ \hline

\textbf{Movement Time}
& \begin{tabular}[c]{@{}c@{}}
\bm{$F(1,23)=4.46,\ p=0.046$}\\
\bm{$d=-0.50,\ \Delta R^2_m=0.010$}
\end{tabular}
& \begin{tabular}[c]{@{}c@{}}
\bm{$F(1,518)=137.57,\ p<0.001$}\\
\bm{$\Delta R^2_m=0.030$}
\end{tabular}
& \begin{tabular}[c]{@{}c@{}}
\bm{$F(2,518)=316.55,\ p<0.001$}\\
\bm{$\Delta R^2_m=0.133$}
\end{tabular} \\ \hline

\textbf{Error Rate}
& \begin{tabular}[c]{@{}c@{}}
$F(1,23)=0.41,\ p=0.530$\\
$d=-0.07,\ \Delta R^2_m=0.003$
\end{tabular}
& \begin{tabular}[c]{@{}c@{}}
\bm{$F(1,518)=33.83,\ p<0.001$}\\
\bm{$\Delta R^2_m=0.023$}
\end{tabular}
& \begin{tabular}[c]{@{}c@{}}
\bm{$F(2,518)=237.05,\ p<0.001$}\\
\bm{$\Delta R^2_m=0.310$}
\end{tabular} \\ \hline

\textbf{Throughput}
& \begin{tabular}[c]{@{}c@{}}
\bm{$F(1,23)=6.18,\ p=0.021$}\\
\bm{$d=0.40,\ \Delta R^2_m=0.011$}
\end{tabular}
& \begin{tabular}[c]{@{}c@{}}
\bm{$F(1,518)=182.80,\ p<0.001$}\\
\bm{$\Delta R^2_m=0.061$}
\end{tabular}
& \begin{tabular}[c]{@{}c@{}}
\bm{$F(2,518)=163.31,\ p<0.001$}\\
\bm{$\Delta R^2_m=0.107$}
\end{tabular} \\ \hline

\rowcolor{gray!15}
Re-entry
& \begin{tabular}[c]{@{}c@{}}
$F(1,23)=3.18,\ p=0.088$\\
$d=-0.31,\ \Delta R^2_m=0.014$
\end{tabular}
& \begin{tabular}[c]{@{}c@{}}
\bm{$F(1,518)=19.63,\ p<0.001$}\\
\bm{$\Delta R^2_m=0.015$}
\end{tabular}
& \begin{tabular}[c]{@{}c@{}}
\bm{$F(2,518)=60.18,\ p<0.001$}\\
\bm{$\Delta R^2_m=0.079$}
\end{tabular} \\ \hline

\rowcolor{gray!15}
Speed
& \begin{tabular}[c]{@{}c@{}}
$F(1,23)=1.00,\ p=0.327$\\
$d=-0.15,\ \Delta R^2_m=0.004$
\end{tabular}
& \begin{tabular}[c]{@{}c@{}}
\bm{$F(1,518)=991.76,\ p<0.001$}\\
\bm{$\Delta R^2_m=0.487$}
\end{tabular}
& \begin{tabular}[c]{@{}c@{}}
$F(2,518)=2.04,\ p=0.132$\\
$\Delta R^2_m=0.004$
\end{tabular} \\ \hline

\rowcolor{gray!15}
Ballistic Time
& \begin{tabular}[c]{@{}c@{}}
$F(1,23)=3.92,\ p=0.060$\\
$d=-0.33,\ \Delta R^2_m=0.011$
\end{tabular}
& \begin{tabular}[c]{@{}c@{}}
\bm{$F(1,518)=14.09,\ p<0.001$}\\
\bm{$\Delta R^2_m=0.012$}
\end{tabular}
& \begin{tabular}[c]{@{}c@{}}
\bm{$F(2,518)=73.75,\ p<0.001$}\\
\bm{$\Delta R^2_m=0.064$}
\end{tabular} \\ \hline

\rowcolor{gray!15}
Correction Time
& \begin{tabular}[c]{@{}c@{}}
$F(1,23)=2.49,\ p=0.128$\\
$d=-0.25,\ \Delta R^2_m=0.008$
\end{tabular}
& \begin{tabular}[c]{@{}c@{}}
\bm{$F(1,518)=223.66,\ p<0.001$}\\
\bm{$\Delta R^2_m=0.136$}
\end{tabular}
& \begin{tabular}[c]{@{}c@{}}
\bm{$F(2,518)=121.12,\ p<0.001$}\\
\bm{$\Delta R^2_m=0.147$}
\end{tabular} \\ \hline

\rowcolor{gray!15}
Correction Distance
& \begin{tabular}[c]{@{}c@{}}
$F(1,23)=2.96,\ p=0.098$\\
$d=-0.08,\ \Delta R^2_m<0.001$
\end{tabular}
& \begin{tabular}[c]{@{}c@{}}
\bm{$F(1,518)=24.21,\ p<0.001$}\\
\bm{$\Delta R^2_m=0.046$}
\end{tabular}
& \begin{tabular}[c]{@{}c@{}}
\bm{$F(2,518)=31.06,\ p<0.001$}\\
\bm{$\Delta R^2_m=0.042$}
\end{tabular} \\ \hline

\end{tabular}%
}
\end{table*}

\begin{figure*}[t!]
  \centering
\subfigure[][\textbf{Movement time}]{%
    \centering
    \includegraphics[width=0.235\textwidth]{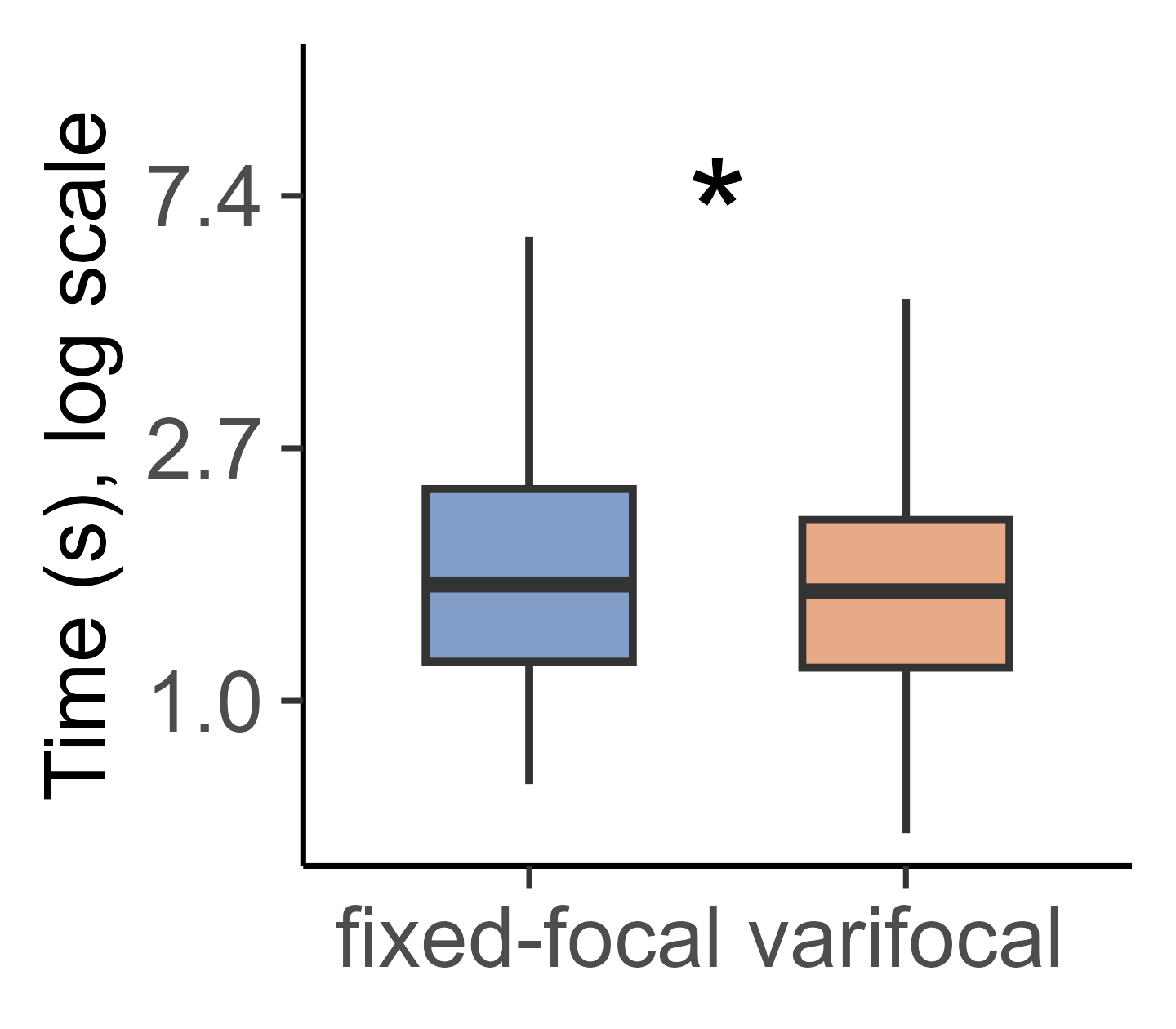}
    \label{fig:StrategyTime}
}
\subfigure[][\textbf{Error rate}]{%
    \centering
    \includegraphics[width=0.235\textwidth]{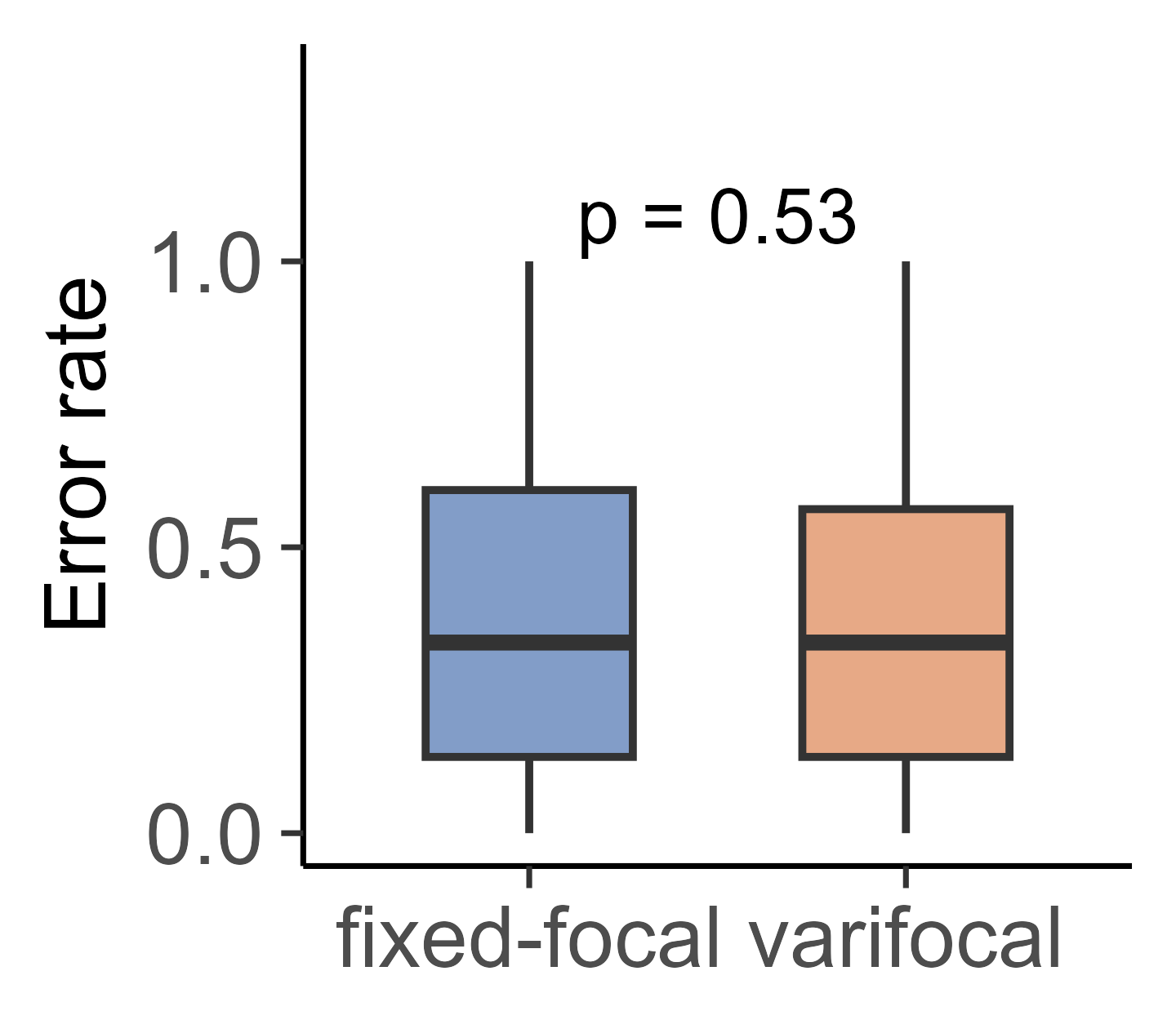}
    \label{fig:StrategyError}
}  
\subfigure[][\textbf{Throughput}]{%
    \centering
    \includegraphics[width=0.235\textwidth]{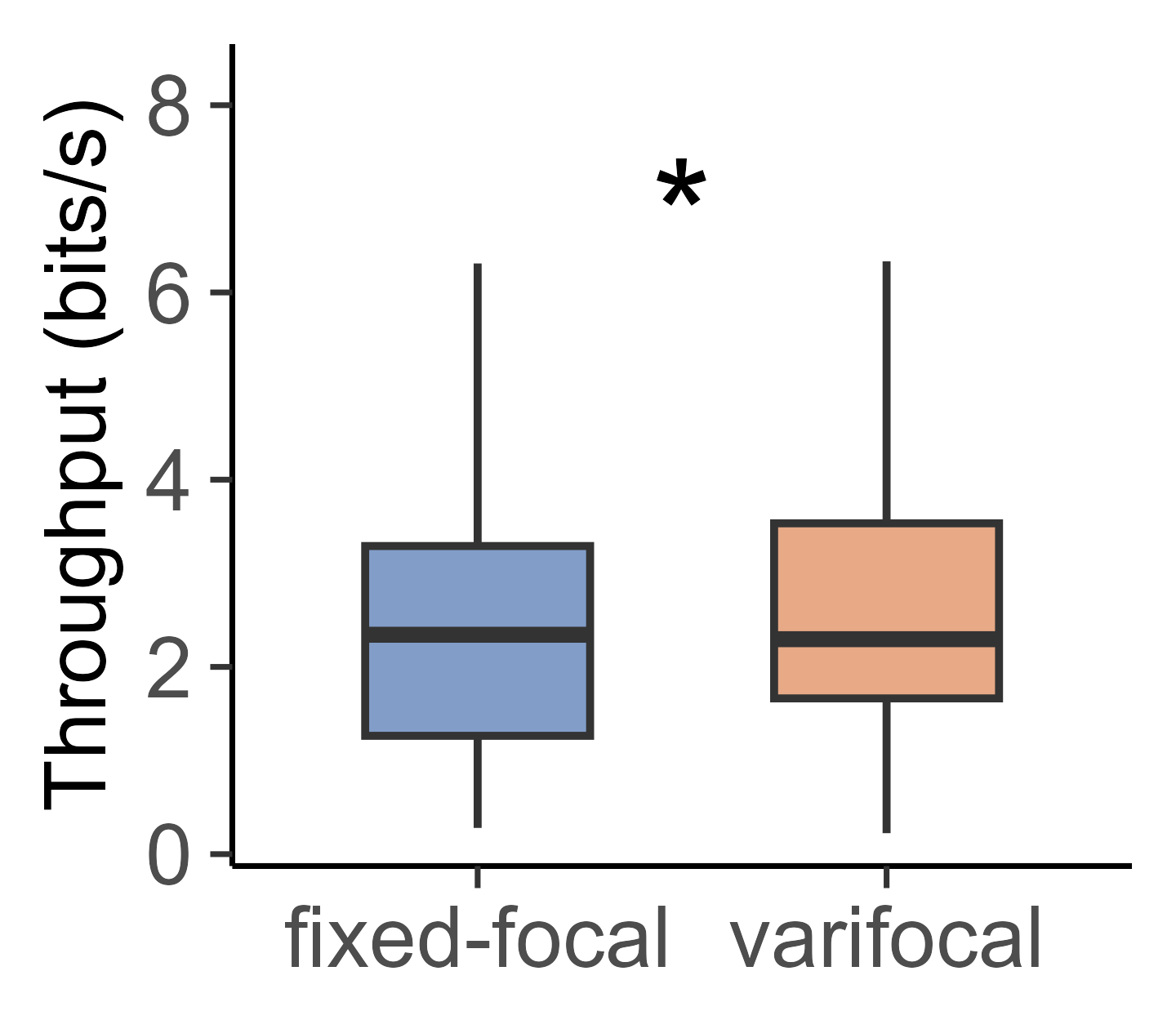}
    \label{fig:StrategyTHP}
}
\subfigure[][Re-entry]{%
    \centering
    \includegraphics[width=0.235\textwidth]{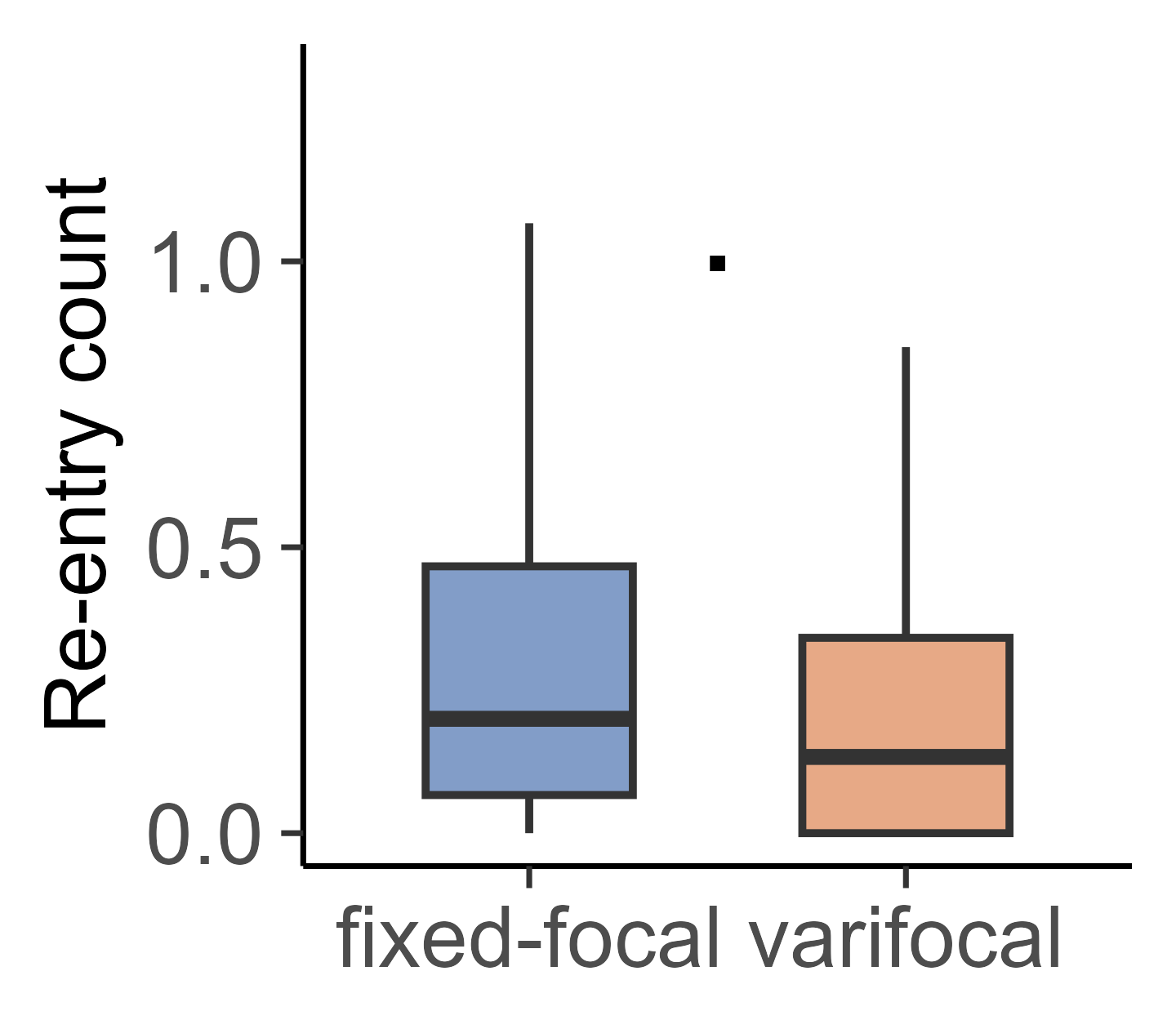}
    \label{fig:StrategySDX}
}
\subfigure[][Speed]{%
    \centering
    \includegraphics[width=0.235\textwidth]{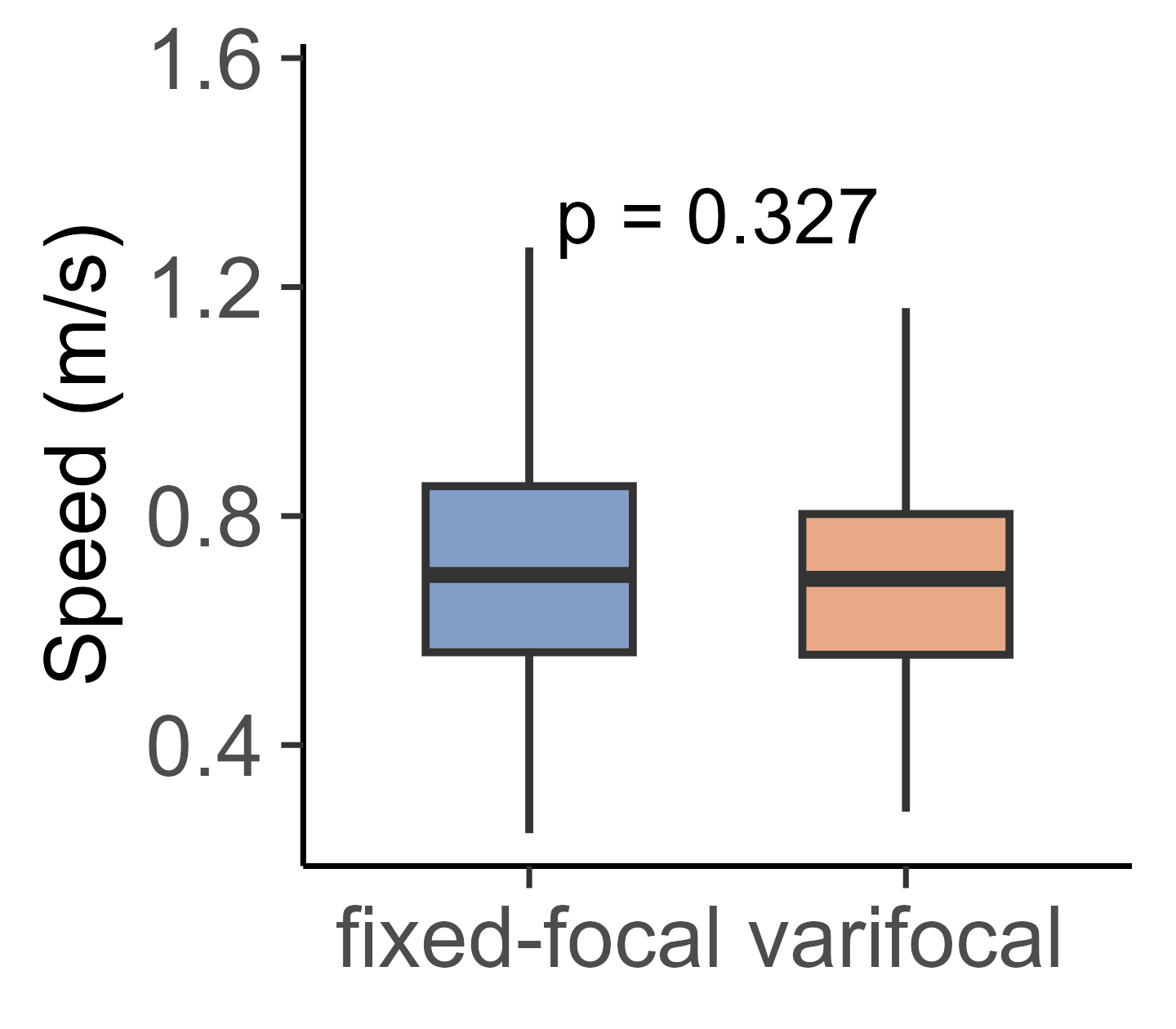}
    \label{fig:SelectionTime}
}
\subfigure[][Ballistic time]{%
    \centering
    \includegraphics[width=0.235\textwidth]{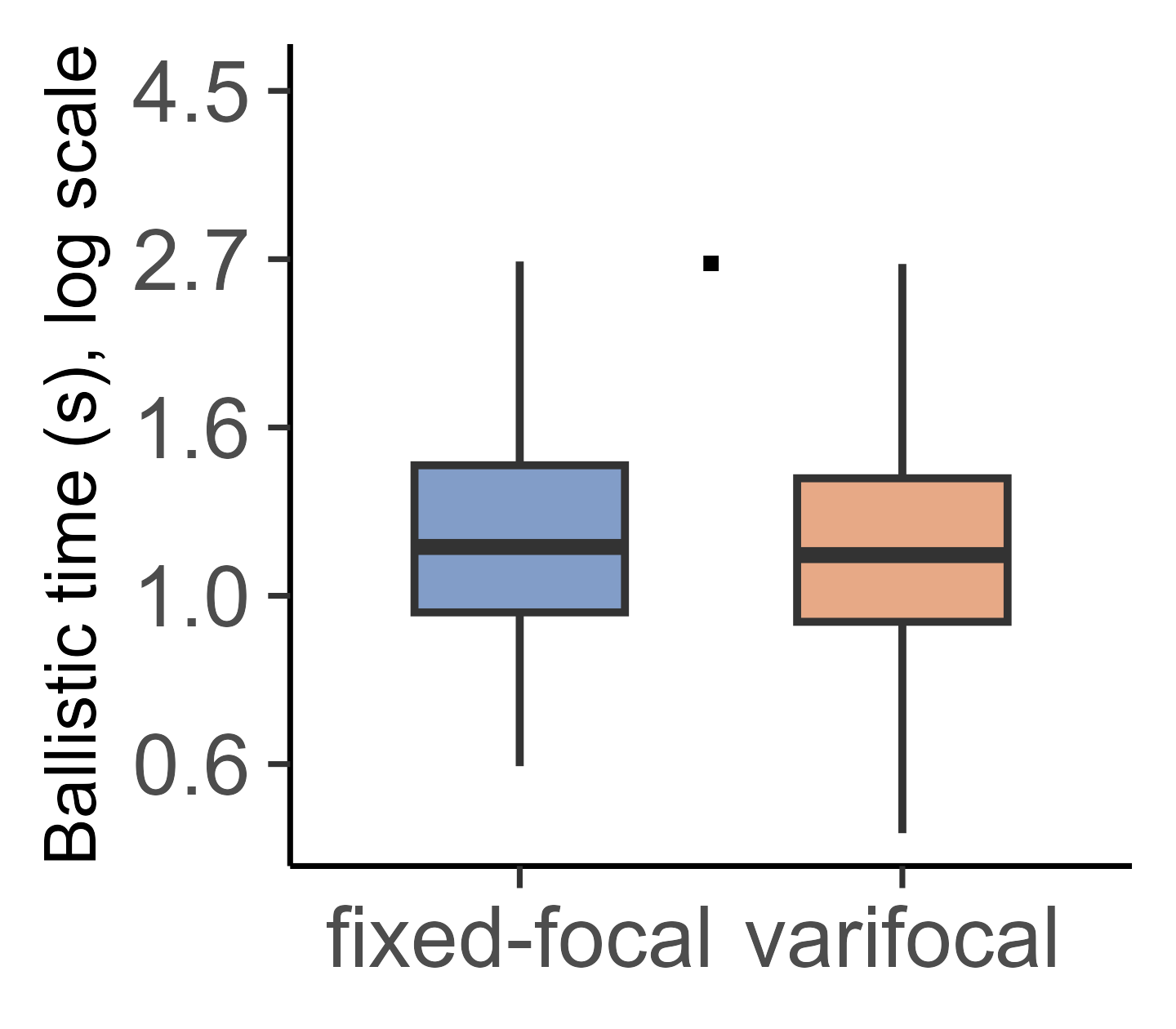}
    \label{fig:SelectionError}
}  
\subfigure[][Correction time]{%
    \centering
    \includegraphics[width=0.235\textwidth]{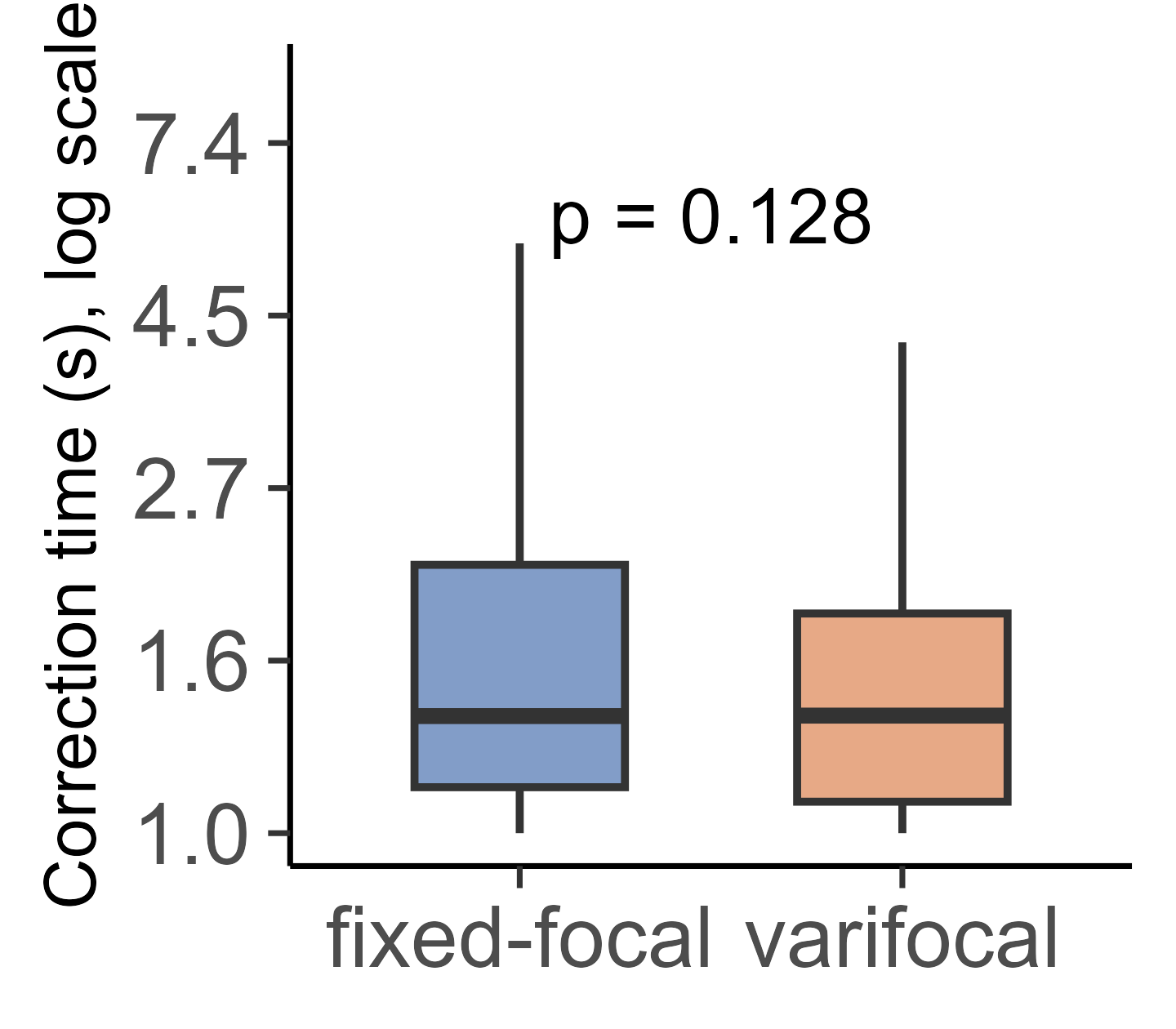}
    \label{fig:SelectionTHP}
}  
\subfigure[][Correction distance]{%
    \centering
    \includegraphics[width=0.235\textwidth]{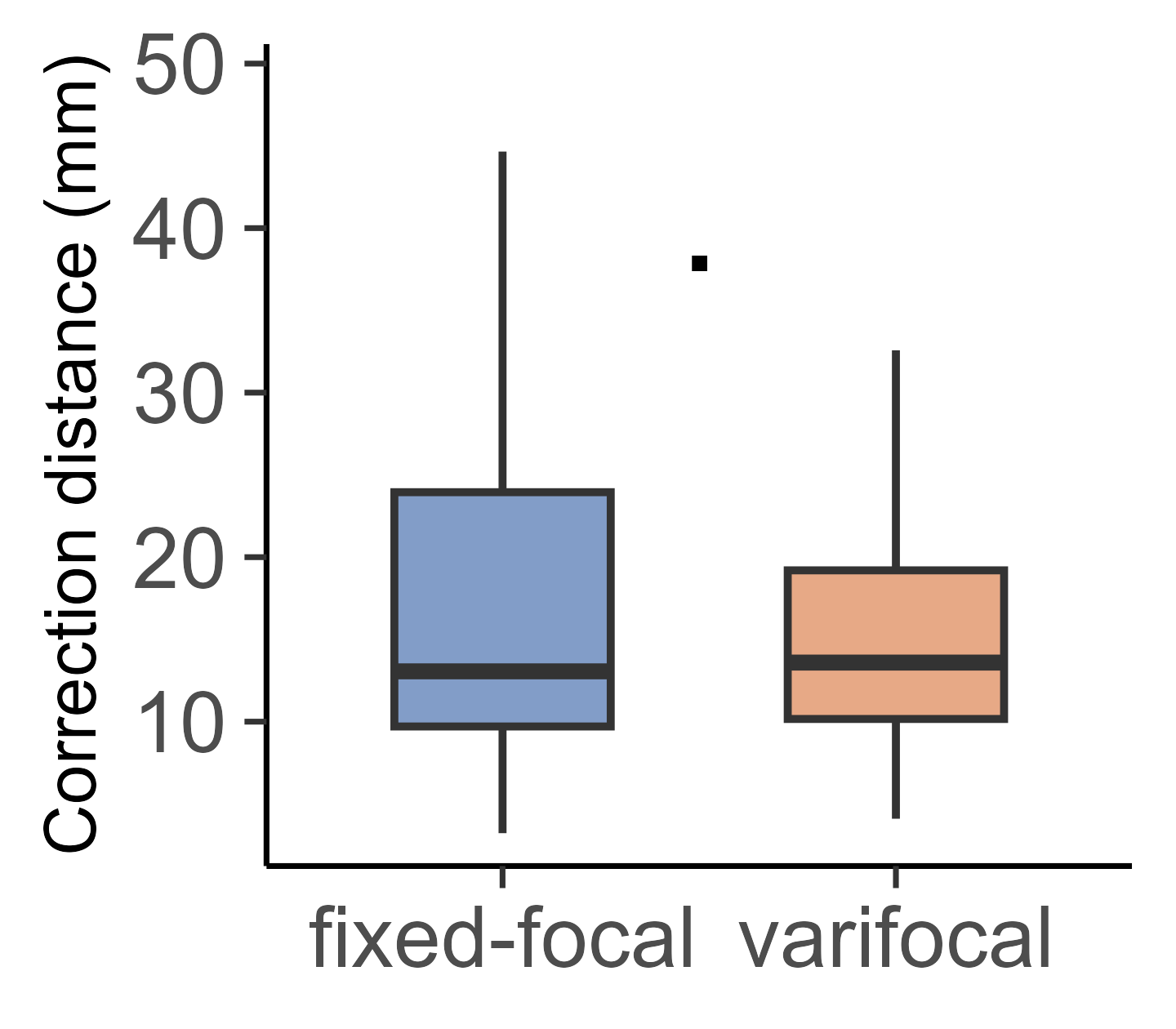}
    \label{fig:SelectionSDX}
}


\caption{Effects of display mode on task performance. Panels (a–h) compare fixed-focal and varifocal conditions for all dependent measures. Time-related measures (a,f,g) are log-transformed, with axes shown in seconds. Display effects are estimated using LMM and visualized as EMM averaged over axis and ID. Significance is indicated by p $<$ 0.05 (*), 0.05 $\leq$ p $<$ 0.1 (.), or exact p values when not significant.}
\label{fig:RMANOVA}
\end{figure*}

\begin{figure}[htbp!]
\centering
\includegraphics[width=.6\linewidth]{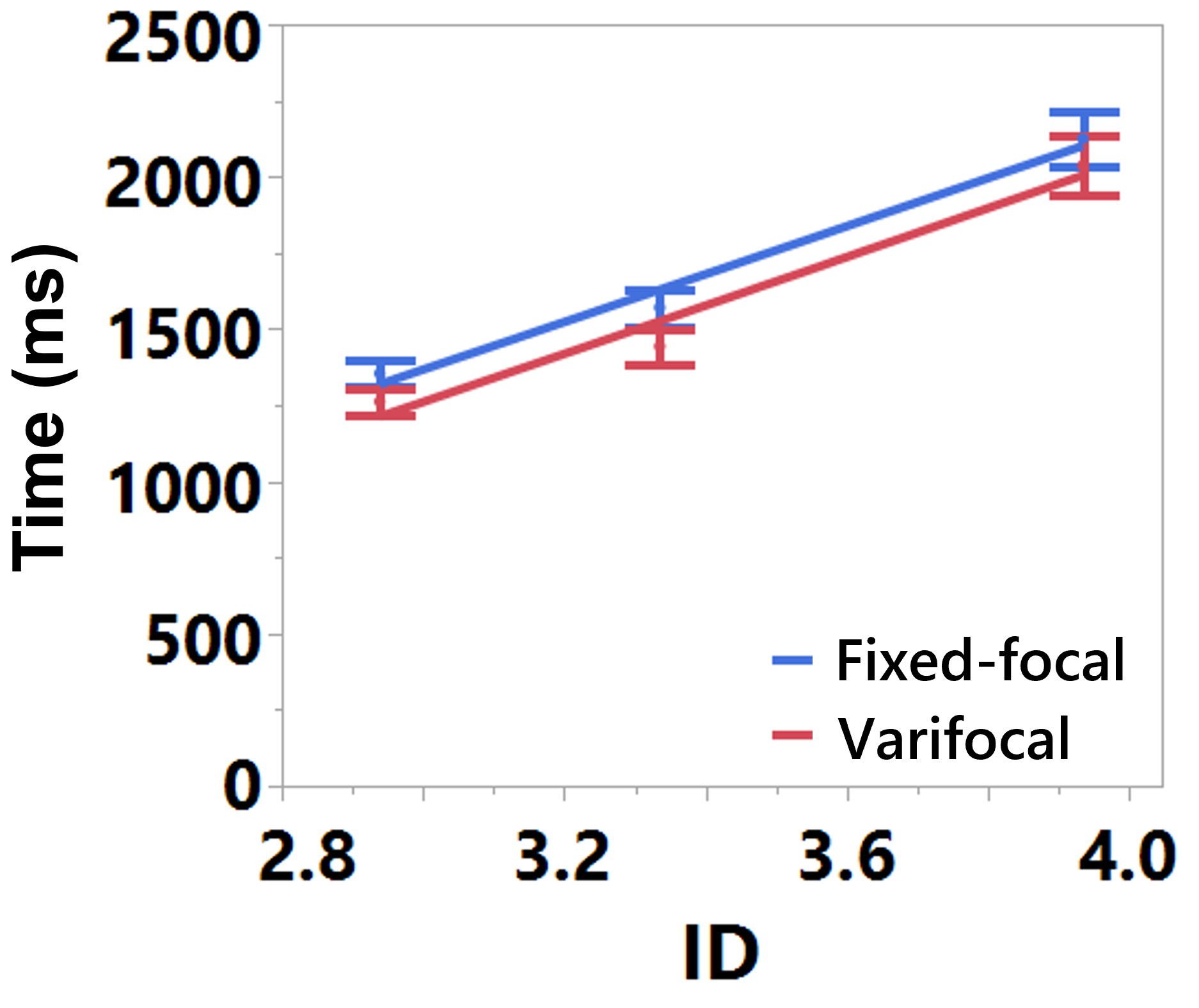}
\caption{Fitts' law results for varifocal and fixed-focal displays.}
\label{fig:fitts}
\end{figure}

\subsection{Overall Display Effects}

When performed for movement direction and ID, the mixed-effects analysis for Study~1 revealed consistent performance advantages for the varifocal display across multiple task metrics.

\additional{The primary performance measures considered in this study were \textbf{movement time}, \textbf{error rate}, and \textbf{throughput}, which are standard outcome measures in Fitts' law evaluations. Additional movement-related metrics were analyzed as secondary measures to provide further insight into the mechanisms underlying any observed performance differences.}

Significant main effects of display mode were observed for \textbf{movement time} and \textbf{throughput}, indicating faster task completion and higher information throughput with the varifocal condition compared to the fixed-focal baseline.

\additional{No significant display effect was observed for \textbf{error rate}, indicating that the performance improvements in movement time and throughput were not achieved at the expense of reduced accuracy.}

\additional{As a secondary analysis, we additionally examined movement-related metrics to better understand the mechanisms underlying the observed performance differences. Several measures exhibited display-related trends in the same direction as the primary results. In particular, \textbf{re-entry count} and \textbf{ballistic time} tended to be lower with the varifocal display, although these differences did not reach statistical significance under the full mixed-effects model. For the remaining measures, including \textbf{speed}, \textbf{correction time}, and \textbf{correction distance}, no statistically significant main effects of display mode were found. Nevertheless, EMM consistently showed numerically improved performance in the varifocal condition across these metrics.}

Across all dependent variables, strong effects of movement direction and ID were observed, reflecting expected influences of task structure.

\additional{We additionally examined the display $\times$ movement direction interaction, which tests whether the effect of display mode differed between lateral and depth movements. This interaction was significant for Movement Time: F(1,518)=4.02, p=0.045, $\Delta R^2_m$=0.0011, and at the threshold of significance for Throughput: F(1,518)=3.87, p=0.050, $\Delta R^2_m$=0.0014. Ballistic time showed a comparable trend, F(1,518)=3.74, p=0.054, $\Delta R^2_m$=0.0036. Other dependent measures did not show reliable display $\times$ movement direction interactions.}

\additional{These interaction patterns indicate that the performance benefit of the varifocal display was more pronounced for depth movements than for lateral movements, consistent with the expectation that VAC-related effects are stronger during depth-oriented interaction.}

In Study~2, we first applied the same mixed-effects specification as in Study~1, including a participant-specific random slope for display, to account for individual differences in responses to the display manipulation. Under this primary specification, the display term did not reach statistical significance for any outcome. Because Study~2 was designed as an ergonomic validation rather than a fully powered confirmatory replication of Study~1, we additionally examined whether the direction of the display-related effects remained consistent under the revised apparatus configuration using a simplified random-intercept model ($1~\mid~participant$). Under this specification, the display-related effects for the primary performance measures remained directionally consistent with Study~1. Participants in Study~2 also showed generally improved overall task performance compared to Study~1, including shorter movement times (approximately 1923/1735 ms in Study~1 vs. 1039/964 ms in Study~2 for fixed/varifocal conditions) and higher throughput values (approximately 2.43/2.67 bps vs. 3.19/3.44 bps), consistent with the reduced biomechanical constraints of the revised setup.

To directly compare the two studies, we additionally fit combined mixed-effects models including study as a factor. These analyses revealed no reliable study $\times$ display interaction for the primary performance measures (Movement Time: $p=0.766$; Throughput: $p=0.961$), suggesting that the magnitude and direction of the display effect did not significantly differ between the two apparatus configurations.
\additional{Together, these results suggest that the display-related effects observed in Study~1 cannot be explained solely by the constraints of the original apparatus.}

\subsection{Fitts' Law Results}
\label{sec:fitts}
We also confirmed that MTs follow Fitts’ Law (\autoref{fig:fitts}), as there is a linear relationship between IDs and Time for varifocal (MT = -992.0431 + 786.37929$*$ID, $R^2$ = 0.96) and fixed-focal displays (MT = -1127.118 + 795.9718$*$ID, $R^2$ = 0.98). 

\begin{figure}[t!]
  \centering
\subfigure[][Time]{%
    \centering
    \includegraphics[width=0.225\textwidth]{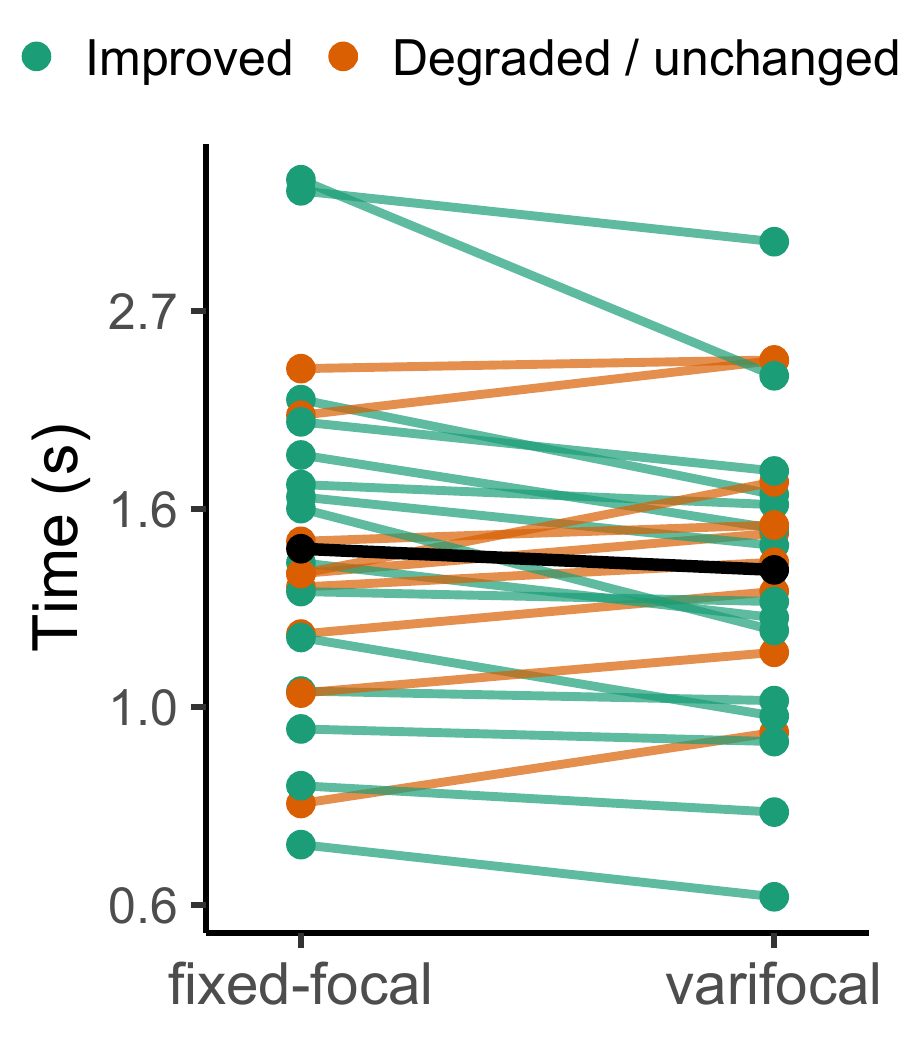}
    \label{fig:slope_time}
}
\subfigure[][Throughput]{%
    \centering
    \includegraphics[width=0.225\textwidth]{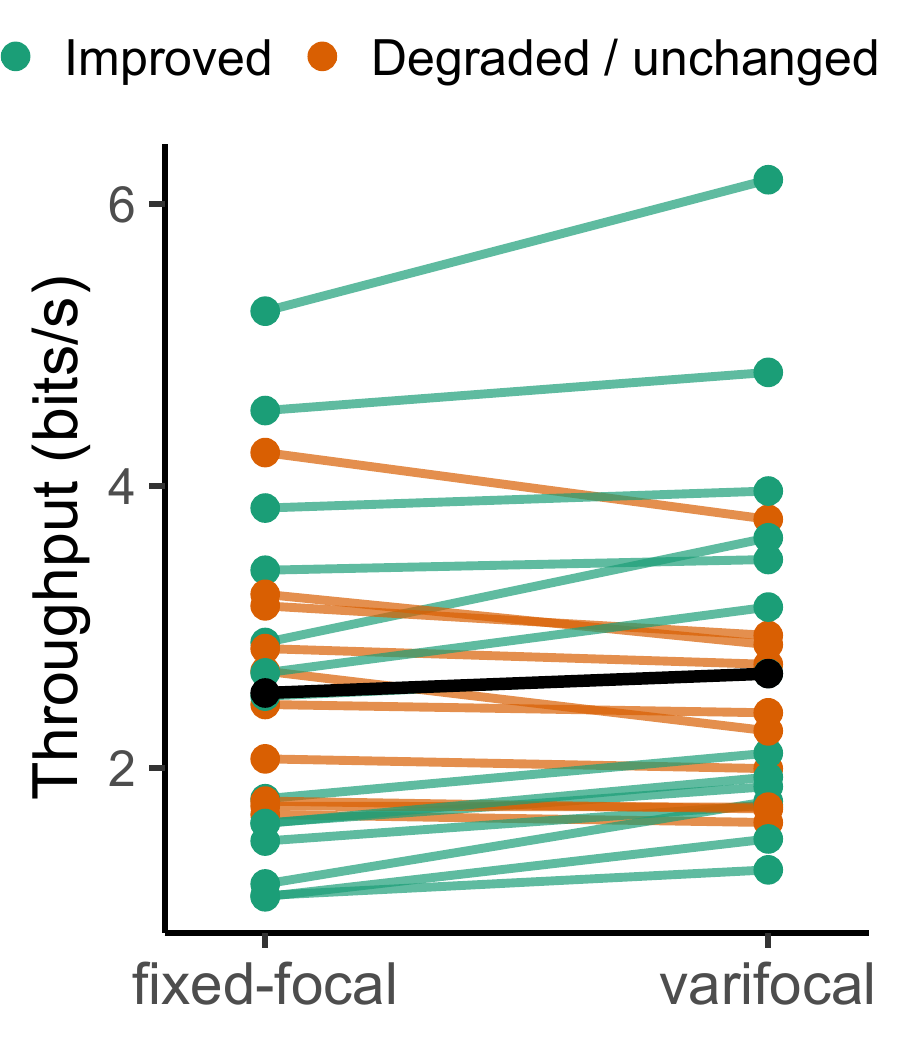}
    \label{fig:slope_thp}
}  
\caption{Participant-specific conditional LMM predictions under fixed-focal and varifocal viewing. Predictions include participant-level random intercepts and random slopes for display. The black line shows the fixed-effects prediction. Colors denote improved vs degraded/unchanged under varifocal viewing, with the direction defined by the metric.}
  \label{fig:slopes}
\end{figure}

\subsection{Baseline-dependent effects of varifocal condition}
\label{sec:skill}
Beyond the average display effects reported above, the mixed-effects analysis revealed substantial inter-individual variability in participants' responses to the varifocal display.

\autoref{fig:slopes} visualizes participant-specific conditional predictions under fixed-focal and varifocal conditions, obtained by combining the fixed effects with participant-level random intercepts and random slopes for display. While the population-level trend indicates improved performance with varifocal viewing on average, individual response patterns vary markedly across participants, including cases of negligible change or performance degradation.

\additional{In these models, the participant-level random intercept represents each participant's baseline performance under the reference display condition (i.e., the fixed-focal condition), while the random slope for display reflects how strongly that participant's performance changes when switching between fixed-focal and varifocal viewing.}
\additional{Because the directionality of the dependent measures differs across metrics, the interpretation of improvement also differs accordingly. For Movement Time, Error Rate, Re-entry Count, Ballistic Time, Correction Time, and Correction Distance, lower values indicate better performance, whereas for Throughput and Speed, higher values indicate better performance.}

To further characterize the structure of these individual differences, we examined the relationship between participant-level random intercepts and display-related random slopes extracted from the LMMs. Across all dependent variables, we observed consistently negative correlations between random intercepts and display slopes (\autoref{tab:intercept_slope_corr}). Notably, several metrics, including Re-entry, Speed, Correction Time, and Correction Distance, exhibited particularly strong negative correlations ($r<-0.7$). Together, these results indicate that \textbf{participants with better baseline performance under the fixed-focal condition tended to exhibit smaller improvements, or even degraded responses, when switching to the varifocal display}. A more detailed analysis of this baseline dependence is presented in Section~\ref{sec:indiff} and \ref{sec:boundary}.

\begin{table}[!h]
\centering
\caption{Correlation between participant-level random intercepts and display-related random slopes extracted from the LMM. Negative correlations indicate that participants with better baseline performance tended to show smaller or negative display-induced changes.}
\label{tab:intercept_slope_corr}
\begin{tabular}{l c}
\hline
\textbf{Variable} & \textbf{Intercept--Slope Correlation} \\
\hline
Movement Time      & $-0.447$ \\
Error Rate         & $-0.572$ \\
Throughput         & $-0.026$ \\
Re-entry Count     & $-0.864$ \\
Speed              & $-0.899$ \\
Ballistic Time     & $-0.394$ \\
Correction Time    & $-0.769$ \\
Correction Distance& $-0.858$ \\
\hline
\end{tabular}
\end{table}

\subsection{Subjective Feedback}

\begin{figure}[t]
\centering
\includegraphics[width=.9\linewidth]{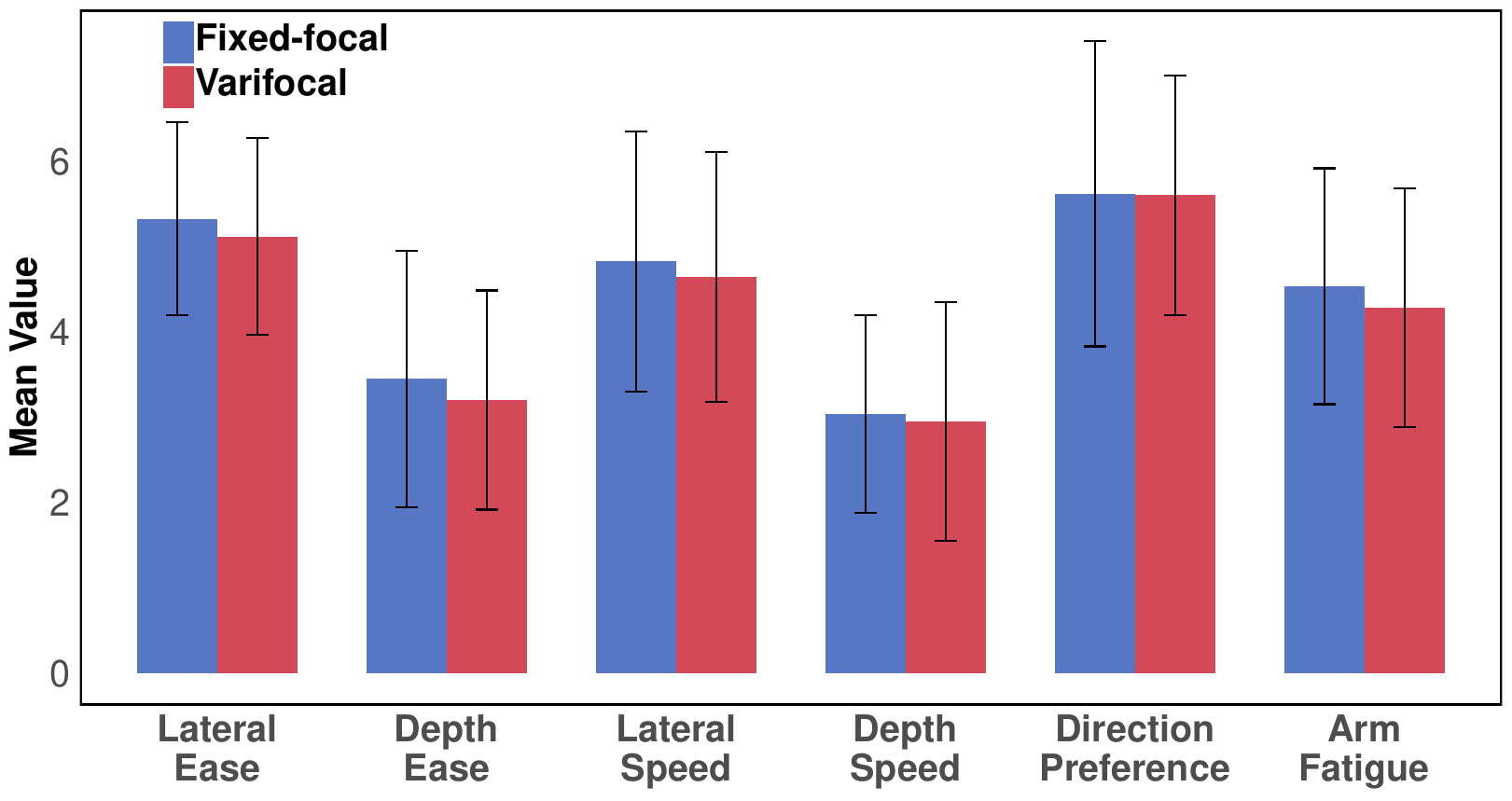}
\caption{Questionnaire results comparing the varifocal and fixed-focal display conditions. All items were rated on a 7-point Likert scale. Lower scores indicate easier, faster, and less tiring interactions. For direction preference, lower scores indicate a preference for lateral interaction, while higher scores indicate a preference for depth interaction.}
\label{fig:report}
\end{figure}

After each display condition, participants completed a questionnaire assessing perceived ease and speed of interaction in lateral and depth directions, interaction preference, and arm fatigue. \autoref{fig:report} summarizes the subjective ratings for the two display conditions.

Overall, the questionnaire did not reveal clear differences between the fixed-focal and varifocal displays. For perceived ease and speed of interaction, participants consistently rated depth interactions as easier and faster than lateral interactions, regardless of display condition. No significant effects of display condition or display-by-direction interactions were observed. Similarly, no notable differences were found between display conditions for interaction preference or reported arm fatigue.

Participants were also asked to report eye discomfort. Under the fixed-focal condition, six participants reported symptoms (three cybersickness and three eyestrain), whereas under the varifocal condition, three participants reported eyestrain. In all cases, reported discomfort was mild and did not prevent task completion.

At the end of the experiment, nine participants reported being able to distinguish between the two display conditions. When asked about overall preference, participants showed a slight tendency toward the varifocal condition (Mean = 3.29, SD = 1.19 on a 7-point scale), although preferences varied across individuals.

\section{Discussion and Limitations}
\subsection{Overview of Hypothesis Evaluation}
\label{sec:hypo_eva}

In this study, we investigated how varifocal stereo displays influence 3D target selection performance and hand movement behavior in peri-personal space. Two hypotheses were formulated: \textit{H1}, which concerns whether varifocal support improves overall interaction performance compared to fixed-focal displays, and \textit{H2}, which examines whether varifocal viewing alters the structure of hand movements during 3D pointing.

Regarding \textit{H1}, our results show that, at the population level, participants achieved significantly better performance under the varifocal condition. Mixed-effects modeling and EMMs revealed reduced movement time and increased throughput compared to the fixed-focal condition. These findings are consistent with prior work suggesting that mitigating the VAC can benefit visually guided interaction, and indicate that varifocal support provides a measurable performance advantage on average.

We further confirmed that task performance followed Fitts’ Law across both display conditions, with movement time increasing monotonically with task difficulty (Section~\ref{sec:fitts}). This pattern held for both lateral and depth-oriented movements, indicating that the experimental task elicited stable speed–accuracy trade-offs and that the observed display effects are not an artifact of task design. While consistent differences were observed between lateral and depth movements in both objective performance and subjective preference, these direction-related effects were present under both display conditions and therefore do not confound the comparison between fixed-focal and varifocal viewing.

With respect to \textit{H2}, our analysis provides more limited evidence. Although varifocal viewing did not significantly alter all components of hand movement, we observed some non-significant trends in several movement-related metrics, such as ballistic time and re-entry behavior, which may indicate differences in movement execution with a varifocal display. However, these effects were not consistent across all participants, indicating that the influence of varifocal displays on movement structure is not uniform.

Together, these results suggest that while varifocal displays yield a modest but statistically detectable average performance benefit(\textit{H1}), the underlying movement mechanisms captured by \textit{H2} are strongly modulated by individual differences. This observation motivates a closer examination of inter-individual variability, which we address in the following section.

\additional{Although the measured interaction-performance improvements were relatively modest, adopting varifocal displays may still be worthwhile for practical AR systems, as reducing the VAC likely also improves visual comfort and reduces eye fatigue during prolonged use, which was not directly evaluated in the present study.}

\additional{
The secondary movement-related measures were included to provide additional insight into the mechanisms underlying the observed performance differences. Although these measures generally showed weaker effects than the primary performance metrics, their overall trends were consistent with the main findings. In particular, reductions in re-entry count and ballistic time are consistent with the possibility that the varifocal display facilitates target acquisition and movement execution, thereby contributing to the improvements observed in movement time and throughput. These results should be interpreted as exploratory and complementary to the primary task-performance outcomes.
}

\subsection{Individual Differences in Response to Varifocal Support}
\label{sec:indiff}
While Section~\ref{sec:hypo_eva} summarizes the average performance benefits of varifocal displays, the mixed-effects results reported in Section~\ref{sec:skill} reveal pronounced inter-individual differences in how participants responded to varifocal support. In particular, both the magnitude and the sign of the varifocal-related display effect varied substantially across participants, with some users benefiting from varifocal viewing and others showing negligible or even negative changes. This variability indicates that population-level averages alone are insufficient to fully characterize user performance.

As shown in Section~\ref{sec:skill}, participant-level random intercepts (reflecting baseline performance under the fixed-focal condition) and display-related random slopes exhibited consistently negative correlations across all dependent variables. This pattern suggests that participants with better baseline performance tended to show smaller improvements, or even degradations, when switching to the varifocal display, whereas participants with poorer baseline performance benefited more from varifocal support.

Importantly, this negative intercept–slope relationship was observed consistently across multiple movement-related metrics, rather than being confined to a single aspect of performance. Such cross-metric consistency suggests that the observed individual differences reflect a structured, baseline-dependent response to varifocal support, rather than measurement noise or isolated task effects. \textbf{To our knowledge, this baseline-dependent interaction pattern has not been previously reported in studies examining varifocal displays for VAC-related 3D interaction.}

Taken together, these findings point to a systematic modulation of varifocal benefits by users’ baseline performance. Users with higher baseline performance under fixed-focal viewing may already compensate for depth-related visual limitations through alternative perceptual or motor strategies, thereby limiting the additional benefit provided by varifocal optics. Conversely, users with lower baseline performance appear to rely more strongly on the additional depth cues afforded by varifocal support.

These results highlight a fundamental challenge in evaluating display technologies that directly interact with human visual perception. Because perceptual sensitivity varies substantially across individuals, performance benefits introduced by a display may not be uniformly expressed at the population level. Prior work has shown that such perceptual individual differences are inherent to visually driven display systems~\cite{hu2024perception}. 
\additional{Such individual differences may also reflect variability in binocular visual function, such as stereoacuity or sensitivity to binocular depth cues, which could influence how strongly the VAC affects performance and therefore how much benefit a user derives from varifocal support.}
Consequently, the absence of a significant average effect does not necessarily imply that meaningful improvements are absent for a subset of users. In the following section, we examine whether these baseline-dependent individual differences could be partially explained by boundary-limited headroom (ceiling/floor effects).

\begin{figure}[t!]
  \centering
\subfigure[][Movement Time]{%
    \centering
    \includegraphics[width=0.225\textwidth]{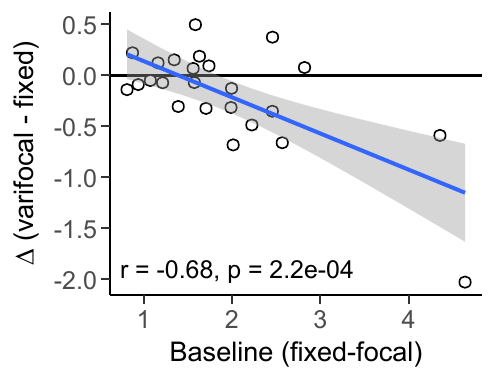}
    \label{fig:delta_time}
}
\subfigure[][Error Rate]{%
    \centering
    \includegraphics[width=0.225\textwidth]{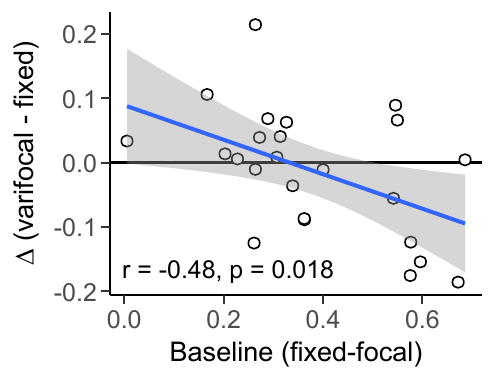}
    \label{fig:delta_ErrorRate}
}  
\subfigure[][Correction Time]{%
    \centering
    \includegraphics[width=0.225\textwidth]{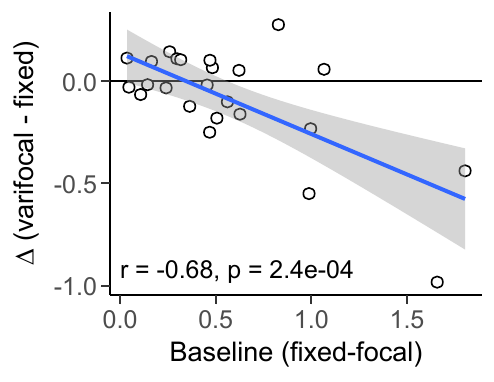}
    \label{fig:delta_ctime}
}  
\subfigure[][Correction Distance]{%
    \centering
    \includegraphics[width=0.225\textwidth]{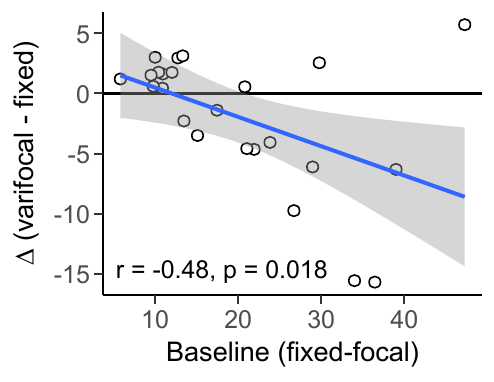}
    \label{fig:delta_cdistance}
} 
\caption{Scatter plots showing the relationship between baseline performance (fixed-focal condition) and display-induced performance change ($\Delta = \bar{y}_{\text{varifocal}} - \bar{y}_{\text{fixed}}$) across participants. Each point represents one participant. The solid line indicates the linear regression fit, and the shaded area denotes the 95\% confidence interval. Subfigures (a–d) correspond to different performance metrics as labeled. \textit{Note: For Movement Time, Error rate, and Correction Time, higher baseline values indicate worse performance. A higher $\Delta$ indicates smaller improvement or even degradation under the varifocal condition.}}
  \label{fig:delta}
\end{figure}

\subsection{Potential Boundary Effects (Ceiling/Floor)}
\label{sec:boundary}
\subsubsection{Assessing Boundary Effects}
The baseline-dependent patterns reported in Section~\ref{sec:skill} raise a natural concern: could the observed relationship between baseline performance and the varifocal benefit be an artifact of boundary-limited headroom (i.e., ceiling or floor effects), rather than reflecting genuine differences in responsiveness across participants? This concern is especially relevant for metrics with hard bounds (e.g., error- or re-entry-related measures), where participants who already perform near the boundary under the fixed-focal condition may have little room to improve, a question that has been previously discussed in Fitts’ law–based interaction studies~\cite{passmore2015fitts}.

To examine this possibility, we conducted a participant-level analysis based on participant-wise mean performance. For each metric, we computed baseline performance as the per-participant mean under the fixed-focal condition, and defined the display-induced change as $\Delta = \bar{y}_{\text{varifocal}} - \bar{y}_{\text{fixed}}$. \autoref{fig:delta} plots $\Delta$ against baseline for four representative metrics (chosen as the primary outcome and the measures most diagnostic of baseline dependence): Movement Time, Error Rate, Correction Time, and Correction Distance. Note that directionality differs by metric. For time- and error-related measures, larger baseline values indicate worse baseline performance. For these measures, a more negative $\Delta$ indicates a larger improvement under varifocal viewing.

If the negative baseline–$\Delta$ relationships were driven solely by boundary effects, we would expect baseline performance to cluster near the corresponding bound for many participants (e.g., near-zero errors or near-minimal times), with $\Delta$ values correspondingly compressed around zero. Our data do not uniformly support this pattern. For Movement Time, baseline values are broadly distributed rather than concentrating near a boundary, yet a significant negative baseline–$\Delta$ trend remains (Figure~\ref{fig:delta_time}). Similarly, Error Rate exhibits a wide spread of baseline values and a consistent negative baseline–$\Delta$ relationship (Figure~\ref{fig:delta_ErrorRate}), suggesting that the observed baseline dependence cannot be attributed solely to a simple saturation effect. Importantly, the strongest and most consistent baseline–$\Delta$ relationships are observed for correction-related measures (Figure~\ref{fig:delta_ctime}, \ref{fig:delta_cdistance}), where baseline performance spans a wide range and varifocal viewing yields larger improvements for participants with poorer baseline performance. These trends align with our interpretation that varifocal viewing particularly reduces the burden during the correction phase of the movement.

\subsubsection{Robustness Check via Influence Diagnostics}
We also observed that the baseline–$\Delta$ relationships for a subset of metrics (notably Speed and Re-entry) could be disproportionately influenced by a small number of participants. We therefore conducted a sanity check using influence diagnostics (Cook’s distance and leverage) and re-estimated the baseline–$\Delta$ correlations after excluding those high-influence participants. Importantly, we did not exclude any participants from the main analysis, because between-participant variability is central to our research question, and even extreme values may reflect genuine individual differences. The influence analysis is reported only to verify that our conclusions are not solely driven by a few participants.

This sanity check shows a clear metric-dependent pattern. The baseline–$\Delta$ relationships remain negative and statistically reliable for the representative measures in \autoref{fig:delta}, particularly for correction-related metrics (Correction Time and Correction Distance) and Error Rate, indicating that these baseline-dependent trends are not attributable to a small number of influential participants. In contrast, the baseline–$\Delta$ associations for Speed and Re-entry become notably weaker after excluding high-influence participants. This sensitivity is expected because these measures are more susceptible to boundary-limited headroom (e.g., floor effects for count-based events) and participant-specific strategies. Consequently, a small number of participants with \emph{unusually} poor baselines can exert a large influence on the estimated correlation. Overall, the robustness of the baseline–$\Delta$ trends in Movement Time and correction-related measures supports our interpretation that boundary effects alone are insufficient to explain the observed individual differences, while also indicating that baseline dependence is not uniform across all metrics.

\subsection{Limitations}

\noindent \textbf{Reduced Visual Realism:}  
Our experimental setup used a plain background without realistic scene elements, which limits the ecological validity of the AR experience. However, this choice aligns with prior controlled studies~\cite{Batmaz:2022:VAC, batmaz2019:DoStereoDeficinciesAffesct, barreramachucaEffectStereoDisplay2019a}, aiming to minimize confounding visual cues such as texture or occlusion. Future work could examine whether the effects observed here hold in visually rich or application-specific AR scenarios.

\noindent \textbf{Lack of Gaze-Contingent Control:}  
Our system does not implement gaze-contingent focal control. While such systems are more representative of natural XR use, we avoided this design to eliminate potential confounds from eye-tracking latency and lens switching delays. Since our task involved rapid gaze shifts between predefined target depths, even minor delays could have affected time-sensitive performance metrics. Instead, we used task-driven focal switching to ensure repeatability and to isolate the effects of the VAC. This limits ecological validity but provides useful benchmarks for future adaptive varifocal designs. \additional{Furthermore, because eye movements were not recorded, we cannot verify whether participants always fixated the currently presented target. If participants instead attended to the cursor or maintained fixation at a different depth, the focal setting would not necessarily match their instantaneous gaze depth, potentially reducing the effectiveness of the varifocal support. Future work should combine eye tracking with direct measurements of vergence and accommodation to better characterize these effects.}

\noindent \textbf{Display Limitations:} 
Our prototype supported a limited range of IPDs, causing calibration difficulties for some participants. The narrow lateral FOV may also have made it harder to perceive targets positioned at the sides of the FOV. These optical limitations could have masked potential interaction effects between movement direction and VAC mitigation. Future studies with wider FOVs and improved optical layout design could help clarify whether depth movements benefit more from a reduced VAC.

\additional{These technical limitations may also have contributed to the relatively low throughput values observed in the present study compared to desktop-based Fitts' law tasks or commercial VR/AR systems. Performing depth-oriented interaction using the current benchtop varifocal AR prototype likely imposed additional visual and ergonomic constraints, including limited FOV optical alignment requirements, and reduced interaction comfort.}

\noindent \textbf{Gender Imbalance:}  
Another limitation is the gender imbalance of our participant sample. While our study focuses on visual perception and eye-hand coordination, where gender differences are not well established, we cannot rule out the possibility that gender may influence interaction behavior. Future studies should include more balanced samples to assess potential gender effects.

\section{Conclusion}
This work examined the effects of varifocal viewing on 3D target selection performance in augmented reality using a standardized Fitts’ law task. Across key interaction metrics, varifocal viewing led to significantly better average 3D pointing performance than fixed-focal viewing, demonstrating its advantage for depth-varying interaction tasks.

At the same time, the results revealed substantial inter-individual variability in how users responded to varifocal viewing. The magnitude and direction of the performance changes varied across participants, exhibiting a baseline-dependent pattern rather than a uniform benefit. These findings indicate that while varifocal displays can improve interaction performance on average, their effectiveness depends on user-specific factors, which should be considered in the evaluation and design of future AR interaction systems.

\backmatter


\bmhead{Acknowledgements}

This work was supported by JSPS KAKENHI (22H00539 and 26K21776), Japan, and Snap Inc. We would like to thank Huakun Liu and Xin Wei for assisting with the pilot study. 

\section*{Declarations}


\begin{itemize}
\item Funding: JSPS KAKENHI (22H00539 and 26K21776), Japan
\item Conflict of interest/Competing interests:
The authors declare that they have no competing interests
\item Ethics approval and consent to participate. Participants were informed about the task in advance and provided informed consent in accordance with the Declaration of Helsinki. This research received ethical approval from the Institutional Review Board of the Nara Institute of Science and Technology (NAIST) with approval number 2024-I-2.
\item Consent for publication:
Not applicable.
\item Data availability:
The datasets generated during and/or analysed during the current study are available from the corresponding author on reasonable request.
\item Materials availability:
Materials are available from the corresponding author on reasonable request.
\item Code availability:
Not applicable.
\item Author contribution:
Conceptualization: All authors; Methodology: All authors; Software: Xiaodan Hu, Mayra Donaji Barrera-Machuca, Anil Ufuk Batmaz; Hardware/Prototype development: Xiaodan Hu, Yan Zhang; Investigation (data collection): Monica Perusquía-Hernández, Xiaodan Hu; Formal analysis: Xiaodan Hu, Anil Ufuk Batmaz; Writing – original draft: All authors; Writing – review \& editing: All authors; Supervision: Wolfgang Stuerzlinger, Kiyoshi Kiyokawa; Funding acquisition: Kiyoshi Kiyokawa.
\end{itemize}

\begin{appendices}




\end{appendices}


\bibliography{sn-bibliography}

\end{document}